\documentstyle[epsfig,eqsecnum,floats,preprint,aps]{revtex}

\tighten
\newcommand{\id}{{\rm I}}

\def\high{\vphantom{\Biggl(}\displaystyle}
\def\balpha{\mbox{\boldmath $\alpha$}}
\def\bbeta{\mbox{\boldmath $\beta$}}
\def\bgamma{\mbox{\boldmath $\gamma$}}
\def\bsigma{\mbox{\boldmath $\sigma$}}
\def\btau{\mbox{\boldmath $\tau$}}
\newcommand{\cosech}[1]{\,\mbox{cosech}\/#1}

\newcommand{\R}{{\bf R}}

\newcommand{\SU}[1]{\mbox{SU}(#1)}

\newcommand{\U}[1]{\mbox{U}(#1)}
\newcommand{\su}[1]{\mbox{SU}(#1)}

\newcommand{\inn}{\in}

\newcommand{\betabf}{\mbox{\boldmath $\beta$}}
\newcommand{\gammabf}{\mbox{\boldmath $\gamma$}}

\newcommand{\sigmabf}{\mbox{\boldmath $\sigma$}}

\begingroup\makeatletter\ifx\SetFigFont\undefined
\def\x#1#2#3#4#5#6#7\relax{\def\x{#1#2#3#4#5#6}}%
\expandafter\x\fmtname xxxxxx\relax \def\y{splain}%
\ifx\x\y   
\gdef\SetFigFont#1#2#3{%
  \ifnum #1<17\tiny\else \ifnum #1<20\small\else
  \ifnum #1<24\normalsize\else \ifnum #1<29\large\else
  \ifnum #1<34\Large\else \ifnum #1<41\LARGE\else
     \huge\fi\fi\fi\fi\fi\fi
  \csname #3\endcsname}%
\else
\gdef\SetFigFont#1#2#3{\begingroup 
  \count@#1\relax \ifnum 25<\count@\count@25\fi
  \def\x{\endgroup\@setsize\SetFigFont{#2pt}}%
  \expandafter\x
    \csname \romannumeral\the\count@ pt\expandafter\endcsname
    \csname @\romannumeral\the\count@ pt\endcsname
  \csname #3\endcsname}%
\fi
\fi\endgroup

\begin{document}

\rightline{CU-TP-1062}
\rightline{hep-th/0207141}
\vskip 1cm
\baselineskip 18pt

\begin{center}
\ \\
\large{{\bf Multicloud solutions with massless and massive monopoles}}
\ \\
\ \\
\normalsize{Conor J. Houghton${}^a$\footnote{Email address: \tt
houghton@maths.tcd.ie} and 
Erick J. Weinberg${}^b$\footnote{Email address: \tt ejw@phys.columbia.edu}}   
\ \\ \medskip
${}^a$\small{\em School of Mathematics, 
Trinity College, Dublin 2,  
Ireland}
\ \\ \smallskip
${}^b$\small{\em Department of Physics, 
Columbia University,  
New York, NY 10027}

\end{center}
\begin{abstract}
\baselineskip=18pt
 Certain spontaneously broken gauge theories contain massless magnetic
monopoles.  These are realized classically as clouds of non-Abelian
fields surrounding one or more massive monopoles.  In order to gain a
better understanding of these clouds, we study BPS solutions with
four massive and six massless monopoles in an SU(6) gauge theory.  We
develop an algebraic procedure, based on the Nahm construction, that
relates these solutions to previously known examples.  Explicit
implementation of this procedure for a number of limiting cases
reveals that the six massless monopoles condense into four distinct
clouds, of two different types.  By analyzing these limiting
solutions, we clarify the correspondence between clouds and massless
monopoles, and infer a set of rules that describe the conditions under
which a finite size cloud can be formed.  Finally, we identify the
parameters entering the general solution and describe their physical
significance.
\end{abstract}

\setcounter{page}{0}
\thispagestyle{empty}
\maketitle
\baselineskip=18pt

\section{Introduction}

Magnetic monopole soliton solutions arise in certain spontaneously
broken gauge theories. After quantization, these give rise to
magnetically charged particles that can be regarded as the
counterparts of the electrically charged elementary particles of the
theory.  Indeed, it is believed that in certain supersymmetric
theories there is an exact duality symmetry \cite{montonen} relating
these two classes of particles.

An interesting new feature arises if the unbroken gauge group 
contains a non-Abelian subgroup.  The massless gauge bosons of this
subgroup transform nontrivially under the gauge group, and thus
carry an ``electric charge''.  Duality then predicts that these should
have massless magnetically-charged counterparts.  These cannot be
realized as isolated classical solutions.  However, evidence for their
existence has been found by analyzing certain multimonopole solutions
\cite{Lee:1996vz}.
These solutions can be viewed as containing a ``cloud'' of non-Abelian
fields that surrounds one or more massive monopoles.  Evidently, this
cloud is the manifestation of the massless monopole.

The previously known solutions of this type either have a single cloud
that generally, although not always, corresponds to a single massless
monopole, or else have two independent clouds.  In this
paper, we obtain a new class of solutions that has a much richer
structure.  By analyzing these, we are able to gain further insight
into the nature of the massless monopoles.

To explain this in more detail, we need to establish some conventions.
Throughout, we consider a gauge theory with an adjoint representation
Higgs field $\Phi$ and restrict ourselves to BPS solutions
\cite{Bogomolny:1975de,Prasad:kr} obeying the Bogomolny equation
\begin{equation}
   D_i \Phi = B_i \equiv  {1\over 2} \epsilon_{ijk}F_{jk} \, .
\end{equation}
Except in this introduction, we will assume that the fields have 
been rescaled so as to set the gauge coupling $e$ to unity.

Both $\Phi$ and the gauge potential $A_i$ can be regarded as elements
of the Lie algebra.  Recall that a basis for this algebra can be
chosen to be a set of $r$ commuting generators $H_i$ that form the
Cartan subalgebra, together with raising and lowering operators
$E_{\balpha}$ associated with the roots.  By an appropriate gauge
transformation, the Higgs expectation value can be chosen to be of the form
\begin{equation}
   \Phi_0 =  {\bf h} \cdot {\bf H} \, .
\label{hdef}
\end{equation}
If the $r$-component vector $\bf h$ has nonzero inner products with
all of the roots, the symmetry breaking is maximal and the unbroken
symmetry group is the maximal torus ${\rm U(1)}^r$. If, however, some
roots are orthogonal to $\bf h$, then there is a non-Abelian unbroken
subgroup $K$ of rank $r'<r$. The roots of $K$ are precisely the roots
that are orthogonal to ${\bf h}$ and the symmetry is broken to
$K\times \U{1}^{r-r'}$.

A basis for the root lattice is given by a set of $r$ simple roots
$\bbeta_a$.  We require that these satisfy ${\bf h} \cdot \bbeta_a \ge
0$.  This determines the set uniquely in the case of maximal symmetry
breaking, but only up to a Weyl transformation when there is
nonmaximal breaking.

At large distances, the magnetic field must commute with the Higgs
field.  Hence, in a direction where the asymptotic Higgs field is of
the form of Eq.~(\ref{hdef}), the magnetic field can be put in the form
\begin{equation}
   B_i = {Q_M \hat r_i\over r^2}  
        + O(r^{-3}) 
\end{equation}
where the magnetic charge $Q_M$ is an element of the Cartan
sub-algebra.  The topological quantization condition requires that
\cite{quantizationRef}
\begin{equation}
   Q_M =  {4 \pi\over e}
     \sum_{a=1}^r k_a {\bbeta_a \over \bbeta_a^2}\cdot{\bf H}
\label{quantizationSolution}
\end{equation}
with the $k_a$ all being integers; for self-dual BPS 
solutions these will all be positive.

In the case of maximal symmetry breaking, the simple roots can be used
to construct a set of $r$ fundamental monopoles.  These are obtained
by embedding the unit SU(2) monopole (appropriately rescaled) in the
SU(2) subgroup associated with each of the $\bbeta_a$.  The
$\bbeta_a$-monopole thus defined has mass $m_a=(4\pi/e^2){\bf h}\cdot
\bbeta_a$ and the radius of its core region is roughly
$1/(e^2m_a)$. It carries one unit of the topological charge $k_a$,
while the remaining $k_b$ all vanish.  Zero-mode analysis shows that
this solution has four zero modes, requiring the introduction of four
collective coordinates.  Three of these specify the position of the
monopole, while the fourth is a U(1) phase; dyonic solutions can be
obtained by allowing this phase to become time-dependent.  An
arbitrary static BPS solution can be interpreted as being composed of
a collection of these fundamental monopoles, with the $k_a$ specifying
the number of each type \cite{Weinberg:1979zt}.  In particular, the
energy is the sum of the component masses while the number of zero
modes is $4 \sum k_a$.

The case of non-Abelian symmetry breaking can be obtained by varying
the Higgs expectation value so that $\bf h$ is orthogonal to some of
the $\bbeta_a$; we will often write $\bgamma_a$ to indicate the
latter.  The BPS mass formula implies that the fundamental monopoles
corresponding to the $\bgamma_a$ should be massless.  Actually, there
is no classical solution corresponding to these monopoles, since the
prescription for embedding the SU(2) monopole gives a trivial vacuum
solution in this limit.  Nevertheless, the formulas for the mass and
counting of zero modes in terms of the $k_a$ remain valid,\footnote{To
maintain the validity of the zero-mode counting, as well as to avoid a
number of pathologies \cite{pathologies} associated with non-Abelian
magnetic charges, we will assume that the total magnetic charge is
Abelian.  This is not a significant restriction, in that any solution
with non-Abelian magnetic charge can be viewed as a purely Abelian one
in which the compensating monopoles are located arbitrarily far away.
} suggesting that the interpretation of higher charged solutions in
terms of component fundamental monopoles should be retained
\cite{Weinberg:ev}.

Some insight can be obtained by considering specific examples.  The
simplest \cite{Weinberg:jh} arises in the context of SO(5) broken to
SU(2)$\times$U(1).  There are two species of fundamental monopoles,
one massive and one massless.  The solutions in which we are
interested contain one of each; we will refer to it as a (1,[1])
solution, with the square brackets indicating that the corresponding
monopole is massless.  Because these solutions turn out to be
spherically symmetric, the BPS equations can be reduced to a set of
ordinary differential equations that can be explicitly solved in terms
of rational and hyperbolic functions.  Examining the solutions, one
finds a massive core of fixed size that is surrounded by a spherical
cloud of radius $a$.  Inside the cloud there is a Coulomb magnetic
field, with both Abelian and non-Abelian components, that corresponds
to the magnetic charge of the massive fundamental monopole.  Outside
the cloud, the non-Abelian components of the magnetic field fall off as
$1/r^3$, leaving the purely Abelian Coulomb field appropriate to the
sum of the two component monopoles.  The energy of the solution is
independent of the cloud radius $a$, which can take on any positive
real value.

In these solutions, the massive monopole is evident and has a well
defined position.  The massless monopole is clearly associated with
the cloud, but it is less clear how to define its position.  To
illustrate this, consider an arbitrary (1,1) solution in the theory
with SO(5) maximally broken to U(1)$\times$U(1).  If the direction of
the Higgs expectation value is varied continuously until the second
fundamental monopole becomes massless, this solution approaches one of
the (1,[1]) solutions.  The separation of the massive monopoles in the
initial solution becomes the cloud parameter in the final solution.
However, the direction of the separation vector has no effect on the
final solution.

More complex solutions have been studied with the aid of the Nahm
construction, which we describe below.  Two examples, both containing
one massless and two massive monopoles, will play an important role in
our considerations.  One is the (1,[1],1) solution \cite{WY} for the
case of SU(4) broken to U(1)$\times$SU(2)$\times$U(1), and the other
is the (2,[1]) Dancer solution \cite{D1,Dancer:kj} for SU(3) broken to
SU(2)$\times$U(1).  In both cases, the massless monopole is manifested
as a cloud that encloses both of the massive monopoles.  Inside the
cloud, one finds the Abelian and non-Abelian Coulomb magnetic fields
appropriate to the charges and positions of the massive monopoles.
Outside the cloud only the Abelian component survives.  As in
the SO(5) example, the cloud is parameterized by a single collective
coordinate that determines its size.  [An Sp(4) solution with one
massless and two massive monopoles is described in
Ref.~\cite{Lee:1997ny}.] 

In order to understand better the nature of these clouds, it would be
helpful to have some examples of solutions with two or more distinct
clouds, or at least with clouds that had more structure.  It might be
expected that such solutions could be obtained simply by adding
additional massless monopoles.  This is not necessarily so.  For
example, the (1,[1],1) solutions in SU(4) can be readily generalized
\cite{WY} to $(1,[1], \dots, [1],1)$ solutions for SU($N$) broken to
U(1)$\times$SU($N-2$)$\times$U(1).  However, although the generalized
solutions have $N-3$ massless monopoles, the spacetime fields display
only a single one-parameter ellipsoidal cloud, no matter what the
value of $N$.  In fact, the solutions are simply embeddings of the
SU(4) solution into the larger group. 

One example with two non-Abelian clouds is the $([1],2,[1])$ solution
in SU(4) broken to U(1)$\times$SU($N-2$)$\times$U(1)
\cite{Ho,HIM}. However, the two clouds do not interact directly and
the structure is no richer than the (2,[1]) Dancer solution.

A more promising choice, on which we will focus in this paper, is the
family of $(2,[2], \dots, [2],2)$ solutions for SU($N$) broken to
U(1)$\times$SU($N-2$)$\times$U(1).  For certain special choices of
parameters, these essentially reduce to a pair of widely
separated $(1,[1], \dots, [1],1)$ solutions.  In general, however,
these solutions are considerably more complex.  At the same time, they
are reasonably tractable.

Let us expand upon this last point.  Except for a few symmetric
solutions, such as the SU(2) unit monopole and the SO(5) solution
$(1,[1])$ solution discussed above, the Bogomolny equation is too
difficult to solve directly.  Instead, it is easier to use a
construction due to Nahm \cite{Nahmconst}.  This Nahm construction has
three distinct steps.  In the first, one solves the Nahm equation,
which is a set of nonlinear ordinary differential equations for a set of
matrices $T_i(s)$.  These Nahm data, as they are called, then specify
a linear differential equation, the construction equation.  Finally,
the spacetime fields are obtained by integration of quantities that
are bilinear in the solutions of the construction equations.

This construction can be carried out explicitly for both the
$(1,[1],1)$ SU(4) solution noted above and the analogous $(1,1,1)$
solution in the maximally broken theory \cite{WY}.  For the $(2,[1])$
Dancer solution, the Nahm equation can be solved, yielding data that
are expressed in terms of elliptic functions.  Except for special
cases, however, the resulting construction equation can only be solved
asymptotically, allowing one to obtain approximations to the spacetime
fields that are valid far from the massive monopole cores.

The remainder of the paper is organized as follows.  In Sec.~II,
we give an overview of the Nahm construction and establish some
conventions.  Next, in Sec.~III, we discuss the parameters entering
the various solutions we will be considering.  We also describe how to
solve the Nahm equation to obtain the data for the SU($N$) $(1,[1],
\dots, [1],1)$ solution, the SU(3) Dancer solution, and, finally, the
$(2,[2], \dots, [2],2)$ solutions that will be our main focus.  In
Sec.~IV we describe how the $(k,[k], \dots, [k],k)$ for SU($N$) can be
obtained algebraically from a knowledge of the
$(k,[k-1],[k-2],\dots,[1])$ solutions of SU($k+1$).  As an example of
this, we show how the $(1,[1], \dots, [1],1)$ solutions can be easily
recovered.  For the case $k=2$, which is our primary interest, we need
some special limits of the Dancer solution; these are described in
Sec.~V.  In Sec.~VI, we use these Dancer results in the construction of
Sec.~IV to obtain $(2,[2], \dots, [2],2)$ solutions for a variety of
limiting cases.  We describe the nature of the clouds in each.  In
Sec.~VII we use these solutions to abstract some general rules
describing how clouds form about massive monopoles. This clarifies the
relationship between clouds and massless monopoles.  Finally, in
Sec.~VIII, we summarize our results and add some concluding remarks.

\section{The Nahm construction}
\label{Nahmsection}

The Bogomolny equation is reciprocal to the Nahm equation. 
This means that the solutions of the Bogomolny equation may be
constructed from the solutions of the Nahm equation and visa
versa. Furthermore, the moduli space of solutions to the Bogomolny
equation is isometric to the moduli space of solutions to the Nahm
equations. The reciprocal relationship is similar to the relationship
between the self-dual Yang-Mills equations and the ADHM data
\cite{ADHM}.

\subsection{Nahm data for $\SU{2}$ monopoles}
The Nahm equation is an equation for a quadruple of Hermitian
matrix functions of a single variable $s$:
\begin{equation}\label{Ne}
\frac{dT_i}{ds}+i[T_0,T_i]=-\frac{i}{2}\epsilon_{ijk}[T_j,T_k].
\end{equation}
Solutions of the Nahm equations are called Nahm data. The size of the
matrices and the boundary conditions that they satisfy determine the
monopole charge in the reciprocal Bogomolny problem.  For an $\SU{2}$
solution with $k$ monopoles of mass $m$, the variable $s$ lies in the
interval $(-m/2,m/2)$ and the data are $k\times k$ matrices.  The
matrices $T_1$, $T_2$ and $T_3$ are required to have simple poles at
$s=\pm m/2$, while $T_0$ must be finite at these end points. By
expanding
\begin{equation}
T_i(s)=-\frac{1}{s\mp m/2}R_i^{\pm}+O(1)
\end{equation}
near $s=\pm m/2$ and substituting back into the Nahm equations, it is
easy to see that the matrix residues $R_i^{\pm}$ satisfy the $\su{2}$
commutation relations 
\begin{equation}
[R^\pm_i,R^\pm_j]=i\epsilon_{ijk}R^\pm_k.
\end{equation}
It is required as a boundary condition that the matrix residues form
an irreducible $k$-dimensional representation of $\su{2}$.  
The case $k=1$ is special, because the one-dimensional representation
of $\su{2}$ is trivial and in fact, for $k=1$ there are no poles at
the end points.  The role played by the boundary condition will become
clear shortly, when we discuss the construction of monopole fields.
Before doing this, it is useful to examine the symmetry groups acting
on the Nahm data.

There is an $\SU{k}$ group action on the data given by
\begin{eqnarray}
T_i&\rightarrow& gT_ig^{-1}\nonumber\\
T_0&\rightarrow& gT_0g^{-1}+i\frac{dg}{ds}g^{-1}
\end{eqnarray}
where $g$ is an $\SU{k}$ function of $s$.  This action does not affect
the monopole fields that are constructed from the Nahm data. This
action is usually referred to as the gauge action. In the $\SU{2}$
case, this action is customarily fixed by requiring $T_0=0$ and
specifying the matrix residues.

There are two further group actions. There is an $\SU{2}$ action which
rotates $(T_1,T_2,T_3)$ as a three-vector:
\begin{equation}
T_i\rightarrow R_{ij}T_j
\end{equation}
and there is a $\R^3$ action which translates the traces:
\begin{equation}
T_i\rightarrow T_i+\lambda_i \id_k.
\end{equation}
These actions correspond to rotation and translation of the monopole
fields themselves. Of course, if the gauge has been fixed by
specifying the residues then the rotation action must include a
compensating gauge transformation.

The spacetime gauge and Higgs fields are obtained from the Nahm data
by first solving the construction
equation,
\begin{equation}\label{coeqn}
\Delta^\dagger v=
\left(-\frac{d}{ds}-T_i\otimes
\sigma_i+\id_k\otimes r_i\sigma_i\right)v(s;{\bf r})=0
\end{equation}
for the $2k$-vector function $v$, which we will refer to as the
construction data.  (To simplify our notation, we will often not 
explicitly show the ${\bf r}$ dependence of $v$.)
An inner product between any two such vector
functions $v$ and $v'$ is given by
\begin{equation}
<v|v'>=\int_{-m/2}^{m/2}v^\dagger v'\, ds.
\end{equation}
With this inner product, there are precisely two linearly
independent normalizable solutions to Eq.~(\ref{coeqn}).  We choose
these to be orthonormal.   

It is easy to see why there are only two solutions.  Near $s=m/2$
\begin{equation}
\Delta^\dagger\approx -\left[ \frac{d}{ds}-\frac{1}{s-m/2}R_i\otimes
\sigma_i\right] 
\end{equation}
where $(R_1,R_2,R_3)$ are irreducible $k\times k$ representation
matrices for $\su{2}$. $R_i\otimes \sigma_i$ has two eigenvalues:
$(k-1)/2$ with degeneracy $k+1$ and $-(k+1)/2$ with degeneracy
$k-1$. This can be shown by choosing an explicit basis and calculating
directly.  However, it is useful to examine the elegant group
theoretical argument, familiar from the addition of
angular momenta, that is given in \cite{H1}. If $(R_1,R_2,R_3)$ is
a representation of $\su{2}$, then $C=\sum R_i^2$ is the Casimir
operator.  For the $k$-dimensional irreducible representation, this is
$C_{\bf k}=\frac{1}{4}(k+1)(k-1)\id_k$.  It is possible to rewrite
$R_i\otimes \sigma_i$ in terms of Casimir operators.  Because ${\bf
k}\otimes{\bf 2}=({\bf k+1})\oplus ({\bf k-1})$,
\begin{equation}
C_{({\bf k+1})\oplus ({\bf k-1})}=(R_i\otimes
\id_2+\id_k\otimes\frac{1}{2}\sigma_i)^2=C_{\bf k}\otimes \id_2+R_i\otimes
\sigma_i+\id_k\otimes C_{\bf 2}
\end{equation}
or,
\begin{eqnarray}
R_i\otimes \sigma_i &=& C_{({\bf k+1})\oplus ({\bf k-1})}-C_{\bf k}\otimes
\id_2- \id_k\otimes C_{\bf 2} \cr 
   &=& C_{({\bf k+1})\oplus ({\bf k-1})}-\frac{(k^2+2)}{4}\id_{2k} \cr
   &=& \frac{(k-1)}{2}\id_{k+1} \oplus \frac{(-k-1)}{2}\id_{k-1} \, .
\end{eqnarray}
Thus, a vector in ${\bf k+1}$ is an
eigenvector of $R_i\otimes \sigma_i$ with eigenvalue $(k-1)/2$ and a
vector in ${\bf k-1}$ is an eigenvector with eigenvalue $-(k+1)/2$.

In order to be normalizable, $v$ must lie in the eigenspace with
eigenvalue $(k-1)/2$ and so it must be perpendicular to the $k-1$
vectors in ${\bf k+1}$. This gives $k-1$ conditions. The pole at
$s=-m/2$ gives another $k-1$ conditions in the same way, and so there
are two normalizable solutions.  If ${v}_1$ and ${v}_2$ are 
an orthonormal basis for these solutions, then the fields are given by
\begin{eqnarray}
\Phi_{ab}&=&<{v}_a|s|{v}_b>=\int_{-m/2}^{m/2}s\,{v}_a^\dagger
\,{v}_b \,ds\nonumber\\ 
(A_i)_{ab}&=&-i<{v}_a|\partial_i|{v}_b>
     =-i\int_{-m/2}^{m/2}{v}_a^\dagger\, \partial_i {v}_b \,ds.
\end{eqnarray}

These fields can be shown to satisfy the Bogomolny equations. This
follows from the fact that the Nahm equations imply that
$\Delta\Delta^\dagger$ commutes with $\id_k\otimes \sigma_i$. Without
dwelling too much on the details, which can be found in \cite{CG}, the
argument runs as follows. By explicit calculation
\begin{equation}
\left(\frac{1}{2}\epsilon_{ijk}F_{jk}\right)_{ab}=
-i\epsilon_{ijk}\int{\int{\partial_jv_a^\dagger(s)
[ \id_{2k}\delta(s-s')-v_c(s) v_c(s')^\dagger]\partial_kv_b(s')\,ds\,ds'}}
\, .  
\end{equation}
Now, the operator $\int [ \id_{2k}\delta(s-s')-v_c(s)
v_c(s')^\dagger]\,ds'$ is a projection operator that projects onto the
null space of $\Delta^\dagger$. This means that it can be replaced by the
projection operator $\int\Delta(\Delta^\dagger\Delta)^{-1}\Delta^\dagger\,ds'$ and so
\begin{equation}
\left(\frac{1}{2}\epsilon_{ijk}F_{jk}\right)_{ab}=
-i\epsilon_{ijk}\int\int\partial_jv_a^\dagger(s) \,
\Delta(\Delta^\dagger\Delta)^{-1}\Delta^\dagger \, \partial_kv_b(s')\,ds'\,ds \, .
\end{equation}
The construction equation implies that $\Delta^\dagger \partial_i v=
-\id_k\otimes\sigma_i v$. Hence, using $[\Delta\Delta^\dagger,\id_k\otimes
\sigma_i]=0$, we obtain
\begin{equation}
\left(\frac{1}{2}\epsilon_{ijk}F_{jk}\right)_{ab}=
2\int\int v_a^\dagger(s) \,\id_k\otimes\sigma_i
(\Delta^\dagger\Delta)^{-1}v_b(s') \,ds'\,ds  \, .
\end{equation}
This is half of the argument. The other half of the argument shows
that $(D_i\Phi)_{ab}$ also equals the right hand side of the above
equation. This half of the argument is very similar to the one just
given, the main difference is that it uses $\Delta^\dagger sv=-v$ as
well as $\Delta^\dagger\partial_iv=-\id_k\otimes \sigma_iv$.

It is also useful to describe how the matrix size
determines the monopole charge. We will follow the proof given in
\cite{H1}. For large $r$, $\Delta^\dagger$ is approximated by
\begin{equation}
\tilde{\Delta}^\dagger=
-\frac{d}{ds}+\left(\frac{1}{s-m/2}+\frac{1}{s+m/2}\right)R_i\otimes\sigma_i+\id_k\otimes x_i\sigma_i  \, . 
\end{equation}
$\tilde{\Delta}^\dagger$ differs from $\Delta^\dagger$ in two
ways. Firstly, the Nahm data have been approximated by their behavior
near the pole. It can be shown that this affects the calculation of
$\Phi$ at order $1/r^2$. The other difference is that the matrix
residues are assumed to be the same at each end.  In fact, while they
are both required to form the same representation, they could differ
by a unitary conjugation. This subtlety is not important since it
corresponds to an $s$-dependent unitary transformation of $v$ and does
not affect the fields.

The eigenvalues of
\begin{equation}
\left(\frac{1}{s-m/2}+ \frac{1}{s+m/2}\right)R_i\otimes\sigma_i+\id_k\otimes
x_i\sigma_i  
\end{equation}
are independent of direction and depend only on $r$. In fact, the
large-$r$ fields calculated using $\tilde{\Delta}$ are spherically
symmetric and so we can take ${\bf r}=(0,0,r)$. This means that
$\id_k\otimes x_i\sigma_i$ commutes with $R_3\otimes
\sigma_3$. Furthermore, $R_3\otimes\sigma_3$ has a unique eigenvector
of eigenvalue $k$ and another of eigenvalue $-k$ lying in the ${\bf
k+1}$ of ${\bf k}\otimes{\bf 2}=({\bf k+1})\oplus({\bf k-1})$. These
eigenvalues are exceptional, in that all the other eigenspaces are
two-dimensional, being spanned by an eigenvector from ${\bf k+1}$ and
an eigenvector from ${\bf k-1}$.  Since the $\pm k$ eigenvectors are in
${\bf k+1}$, we know how they are acted on by $R_i\otimes \sigma_i$:
both of them have eigenvalue $(1-k)/2$. Since $({\bf 1}_k\otimes
x_i\sigma_i)^2=r^2 \id_{2k}$, we also know how they are acted on by $
\id_k\otimes x_i\sigma_i$: they have eigenvalues $r$ and $-r$. Let us
denote the two eigenvectors $v_+$ and $v_-$ respectively.

Now, if we substitute $v(s)=g_{\pm}(s)\,v_\pm$ into the approximate
construction equation $\tilde{\Delta}^\dagger v=0$, we get
\begin{equation}
\frac{dg_\pm}{ds}+\frac{k-1}{2}\left(\frac{1}{s-m/2}+
\frac{1}{s+m/2}\right)g_\pm \, \pm \,  r\,g_\pm=0   \, , 
\end{equation}
which is solved by 
\begin{equation}
g_\pm=\left(s^2-\frac{m^2}{4}\right)^{(k-1)/2}e^{\pm rs}.
\end{equation}
Because
$v_\pm$ are pointwise orthogonal,  we need only consider the
diagonal components of $\tilde{\Phi}$. For large $r$, the exponential
in $g_+$ means that the value of any integral of a polynomial times 
$g_+^2$ is dominated by
the value of the integrand near $m/2$. This allow us to approximate
\begin{equation}
\int_{-m/2}^{m/2}{g_+^2 \,ds}=
\int_{-m/2}^{m/2}{\left(s^2-\frac{m^2}{4}\right)^{k-1}e^{2rs}\,ds}  
\approx e^{mr} \int_{-\infty}^0 e^{2ru} u^{k-1} (m+u)^{k-1} \, du
\end{equation}
and 
\begin{equation}
\int_{-m/2}^{m/2}{s\,g_+^2\,ds}=
\int_{-m/2}^{m/2}{s\left(s^2-\frac{m^2}{4}\right)^{k-1}e^{2rs}\,ds}
\approx e^{mr} \int_{-\infty}^0 e^{2ru} u^{k-1} (m+u)^{k-1}
     \left({m\over 2} + u\right) \, du  \, .
\end{equation}
We can then integrate by parts and show that
\begin{equation}
\tilde{\Phi}_{11}=
\frac{\int_{-m/2}^{m/2}{sg_+^2\,ds}}{\int_{-m/2}^{m/2}{g_+^2\,ds}} =
\frac{m}{2}-\frac{k}{2r} + O(m/r^2)  \, .
\end{equation}
$\tilde{\Phi}_{22}$ works the same way, establishing the relation between the
charge and the size of the Nahm matrices. We will use similar
techniques below to calculate approximate monopole fields.

The construction is remarkably easy to implement in the case of a
single SU(2) monopole, where it even predates the Nahm equation
\cite{N1}.  For $k=1$ the Nahm matrices reduce to numerical functions
of $s$.  The Nahm equations imply that these must be constants, which
turn out to give the position of the monopole.  By translational
invariance, this can be taken to be the origin. Thus
\begin{equation}
v_a(s;{\bf r})=N({\bf r})\, e^{{\bf r}\cdot\sigmabf s}v^0_a({\bf r})
\label{VforUnitMono}
\end{equation}
where the $v_a^0$ are orthonormal and the normalization factor
\begin{equation}
    N = \sqrt{2r}\, e^{-mr/2} \left[1 - e^{-2mr} \right]^{-1/2}
\label{normalizationfactor}
\end{equation}
is obtained by noting that 
\begin{equation}
\int_{-m/2}^{m/2}e^{2{\bf r}\cdot\sigmabf s}ds=\frac{\sinh{mr}}{r}\, \id_2
\end{equation}
Because
\begin{equation}
\int_{-m/2}^{m/2}s\,e^{2{\bf r}\cdot\sigmabf s}ds=
\frac{1}{2r^3}\left(mr\cosh{mr}-\sinh{mr}\right){\bf r}\cdot\sigmabf 
\end{equation}
the Higgs field is 
\begin{equation}
   \Phi_{ab}({\bf r}) = {1\over 2} \left(m \coth{mr} - {1 \over r}\right)
         {v_a^0}^\dagger \,\hat{\bf r}\cdot \bsigma \, v_b^0  \, .
\end{equation}
The usual ``hedgehog gauge'' form of the one-monopole Higgs field is
obtained by taking $v_1^0 = (1, 0)^t$ and $v_2^0 = (0, 1)^t$.  Another
possibility is 
\begin{equation}
    v_1^0 =  \psi({\bf r})  \qquad\qquad
    v_2^0 =  \bar\psi({\bf r}) 
\end{equation}
where the two-component spinors
\begin{equation}
     \psi({\bf r}) = \sqrt{r-z \over 2r}
           \left(\matrix {\high{x-iy \over r-z} \cr \high 1}\right) 
     \qquad\qquad
    \bar\psi({\bf r}) = \sqrt{r-z \over 2r}
    \left(\matrix { \high 1 \cr\high -{x+iy \over r-z}} \right)  \, ,
\label{spinordef}
\end{equation}
which are eigenvectors of $\hat{\bf r}\cdot \bsigma$ with eigenvalues
$\pm 1$, satisfy $\psi^\dagger \psi = \bar \psi^\dagger \bar \psi =1$,
$\psi^\dagger \bar \psi=0$.  With this choice, the Higgs field is
everywhere proportional to $\sigma_3$.  The singularity in $\psi({\bf
r})$ along the positive $z$-axis is the Dirac string that appears
whenever a monopole solution is written with a uniform Higgs field
direction.

\subsection{The Nahm construction for SU($N$)\label{ncogg}}

For $\SU{2}$ monopoles, the boundary values of $s$ are given by the two
eigenvalues $\pm m/2$ of the asymptotic Higgs field $\Phi_0$.  In the
case of $\SU{N}$ monopoles, the asymptotic Higgs field has more than
two eigenvalues and the Nahm data are matrix functions over a
subdivided interval defined by the eigenvalues. In order to describe
this, it is convenient to choose a particular basis for the Cartan
generators.  We choose diagonal generators $H_i$ so that
\begin{equation}
\Phi_0={\bf h}\cdot{\bf H}=\mbox{diag}\,(s_N,s_{N-1},\ldots,s_1)
\label{Phi0eigenvalues}
\end{equation}
where $s_1\le s_2\le \ldots \le s_N$. In the case of maximal symmetry
breaking, none of these eigenvalues are equal and so all the
inequalities are strict. The magnetic charge is 
\begin{equation}
Q_M=4\pi\,\mbox{diag}\,(k_{N-1},k_{N-2}-k_{N-1},\ldots,k_1-k_2,-k_1) \, .
\end{equation}

With this basis of Cartan generators, the $n$th fundamental monopole,
with $k_m=\delta_{nm}$, is obtained by embedding an appropriately
rescaled $\SU{2}$ monopole solution in the $2\times 2$ block at the
intersection of the $(N-n)$th and $(N+1-n)$th rows and columns. This
fundamental monopole then has mass $2\pi(s_{n+1}-s_n)$.

In the Nahm construction, the corresponding Nahm data are defined on an
interval divided into $N-1$ subintervals: $(s_1,s_2)$, $(s_2,s_3)$, and
so on to $(s_{N-1},s_N)$. Each of these subintervals can be thought of
as corresponding to a different fundamental monopole. The size of the
Nahm matrices over that interval is given by the number of monopoles
of that type. Thus, the data reciprocal to a
$(k_1,k_2,\ldots,k_{N-1})$ monopole solution are $k_1\times k_1$ matrices for
$s\inn(s_1,s_2)$, $k_2\times k_2$ matrices for $s\inn(s_2,s_3)$ and so
on to $k_{N-1}\times k_{N-1}$ matrices for $s\inn(s_{N-1},s_N)$. The
length of the subinterval determines the mass of the fundamental
monopole.
When describing Nahm data, it is sometimes useful to use a
skyline diagram: a step function over the interval whose height in a
subinterval is given by the size of the Nahm matrices in that
subinterval.

There are boundary conditions relating the triplets $(T_1,T_2,T_3)$ on
either side of one of the subdivision points $s_n$. (There are no
boundary conditions on $T_0$.)  If
$k_n\not=k_{n+1}$ these boundary conditions are quite simple.  
For the skyline diagram
\begin{equation}
\begin{array}{c}
\begin{picture}(200,90)(60,600)
\thinlines
\put( 60,620){\line( 1, 0){200}}
\put(100,680){\line( 1, 0){ 60}}
\put(160,680){\line( 0, -1){ 20}}
\put(160,660){\line( 1, 0){ 70}}
\multiput(160,660)(0,-8){6}{\line(0,-1){  4}}
\multiput(100,680)(-8,0){3}{\line(-1,0){  4}}
\multiput(230,660)( 8,0){3}{\line( 1,0){  4}}
\put( 95,623){\vector(0,1){  56}}  
\put( 95,623){\vector(0,-1){  0}} 
\put(160,620){\line( 0,-1){  5}}
\put( 81,648){\makebox(0,0)[lb]{\smash{\SetFigFont{12}{14.4}{rm}$k_n$}}}
\put(145,606){\makebox(0,0)[lb]{\smash{\SetFigFont{12}{14.4}{rm}$s=s_n$}}}
\put(240,638){\makebox(0,0)[lb]{\smash{\SetFigFont{12}{14.4}{rm}$k_{n+1}$}}}
\put(235,623){\vector(0,1){  36}}
\put(235,623){\vector(0,-1){  0}}
\end{picture}\end{array}
\end{equation}
\noindent the Nahm triplet, $(T_1,T_2,T_3)$, is
a triplet of $k_n\times k_n$ matrices over the left interval and of
$k_{n+1}\times k_{n+1}$ matrices over the right interval. As $t=s-s_n$
approaches $0-$, it is required that
\begin{equation}
T_i(t)=\left(\begin{array}{cc}R_i/t+O(1) &
      O(t^{(m-1)/2}) \\ O(t^{(m-1)/2}) &
      T_i^\prime+O(t)
    \end{array}\right)\label{frombelow}
\end{equation}
where $m=k_n-k_{n+1}$ and the $k_{n+1}\times k_{n+1}$ matrix
$T_i^{\prime}$ is the nonsingular limit of the right interval Nahm
data at $t=0+$.  The $m\times m$ residue matrices $R_i$ in
(\ref{frombelow}) must form an irreducible $m$-dimensional
representation of $\su{2}$. Since the one-dimensional representation
is trivial, there is no singularity when $m=1$.

Thus, when the data are a different size on each side of a junction
point, the smaller matrices match up continuously with sub-matrices of
the larger matrices.  The vector $v(t;{\bf r})$ in the construction
equation is split in the same way, with $k_n$ components carrying
through the boundary:
\begin{equation}
v(t)=\left(\begin{array}{c}O(t^{(m-1)/2}) \\ v^\prime \end{array}\right)
\end{equation}
where $v^\prime$ is the the nonsingular limit from the right interval.

When $k_1=k_2$ the situation is different. For the situation described
by the skyline diagram
\begin{equation}
\begin{array}{c}
\begin{picture}(200,70)(60,600)
\thinlines
\put( 60,620){\line( 1, 0){200}}
\put(100,660){\line( 1, 0){ 100}}
\multiput(160,660)(0,-8){6}{\line(0,-1){  4}}
\multiput(100,660)(-8,0){3}{\line(-1,0){  4}}
\multiput(200,660)( 8,0){3}{\line( 1,0){  4}}
\put( 95,623){\vector(0,1){  36}}  
\put( 95,623){\vector(0,-1){  0}} 
\put(160,620){\line( 0,-1){  5}}
\put( 81,638){\makebox(0,0)[lb]{\smash{\SetFigFont{12}{14.4}{rm}$k$}}}
\put(145,606){\makebox(0,0)[lb]{\smash{\SetFigFont{12}{14.4}{rm}$t=0$}}}
\end{picture}\end{array}
\end{equation}
the Nahm matrices may be discontinuous across $t=0$. In fact, there are
additional data, called jumping data, associated with the $t=0$
junction. 
Let 
\begin{equation}
\delta T_i=T_i(0-)-T_i(0+)
\end{equation}
be the discontinuity in the Nahm matrices. The jumping data are a 
$k$-vector of 2-component spinors $a_{r\alpha}$ satisfying the jump equation
\begin{equation}\label{jeind}
(\delta T_i)_{rs}= a^\star_{r\alpha}(\sigma_i)_{\alpha\beta}a_{s\beta}.
\end{equation}
The $r$ and $s$ run from one to $k$ and $\alpha$ and $\beta$ are
spinor indices running from one to two.

There are also additional construction data, $S\in{\bf C}$, associated with the junction. 
The construction vector $v(t;{\bf r})$ obeys
\begin{equation}
\delta v=v(0-)-v(0+)=Sa
\end{equation}
where here $a$ is considered a $2k$-vector rather than a $k$-vector of
2-spinors. This reordering of $a$ is done in the obvious way; $v$
could also be considered a $k$-vector of 2-spinors with the Nahm
matrices acting on the vector index and the Pauli matrices acting on
the spinor index.

The extra construction data $S$ enter the inner product and in the
construction of the fields.
Let us write $V=(v_1,v_2,\ldots,v_{N-1};S_{p_1},S_{p_2},\ldots)$ for
the construction data. Here $v_n$ is a solution to the construction
equation in the interval $(s_n,s_{n+1})$, with $v_n$ and $v_{n+1}$
satisfying the appropriate boundary conditions at $s=s_n$. The $S_p$'s
are the jumping data at any point $s_p$ where $k_n=k_{n+1}$. The number of
these zero jumps is certainly less than $N-2$, but depends
on the charge.  In this case,
\begin{equation}\label{genortho}
<V|V'>=\sum_{n=1}^{N-1}\int_{s_n}^{s_{n+1}}{v_n^\dagger
v_n'ds}+\sum_{p} {S^*_pS_p'} 
\end{equation}
where $p$ sums over the points where $k_n=k_{n+1}$.
If $\{{V}_1,\ldots,{V}_N\}$ is an orthonormalized basis for
the construction data then 
\begin{eqnarray}\label{Higgs}
\Phi_{ab}&=&<{V}_a|s|{V}_b>
=\sum_{n=1}^{N-1}\int_{s_n}^{s_{n+1}}{(v_a)_n^\dagger 
s(v_b)_nds}+\sum_{p} {(S_a)^*_ps_p(S_b)_p}\nonumber\\ 
(A_i)_{ab}&=&-i<{V}_a|\partial_i|{V}_b>=
-i\left(\sum_{n=1}^{N-1}\int_{s_n}^{s_{n+1}}{(v_a)_n^\dagger
\partial_i(v_b)_nds}+ \sum_{p} {(S_a)^*_p\partial_i(S_b)_p}\right) \, .
\end{eqnarray}
Of course, we have not shown that there is a $N$-dimensional family
of solutions, or that the fields will have the right charge, or even
that the fields satisfy the Bogomolny equation. In fact, all of these
things are easily demonstrated using the same sort of argument
as in the $\SU{2}$ case described above \cite{Nahmconst,HM}. 

If two $s_i$'s are coincident, there is a zero thickness subinterval
in the Nahm interval. This means that there is a non-Abelian residual
symmetry and some of the monopoles have zero mass.
The boundary conditions for Nahm data in this situation can
be described in terms of those explained above, by formally imagining
the zero thickness subinterval as a limit of a
subinterval of finite thickness. The Nahm data on this subinterval
become irrelevant in the limit, but the height of the skyline on the
vanishing subinterval affects the matching condition between the Nahm
matrices over the subintervals on either side.

The skyline for a (2,1)-monopole, for example, is given by
\begin{equation}
\begin{array}{c}
\begin{picture}(205,85)(55,605)
\thinlines
\put( 60,620){\line( 1, 0){200}}
\put(100,620){\line( 0, 1){ 40}}
\put(100,660){\line( 1, 0){ 60}}
\put(160,660){\line( 0,-1){ 20}}
\put(160,640){\line( 1, 0){ 70}}
\put(230,640){\line( 0,-1){ 20}}
\multiput(160,640)(0.00000,-8.00000){3}{\line( 0,-1){  4.000}}
\put( 95,623){\vector(0,1){  36}}  
\put( 95,623){\vector(0,-1){  0}} 
\put(235,623){\vector(0,1){  15}}
\put(235,623){\vector(0,-1){  0}}
\put(230,620){\line( 0,-1){  5}}
\put(100,620){\line( 0,-1){  5}}
\put(160,620){\line( 0,-1){  5}}
\put( 84,638){\makebox(0,0)[lb]{\smash{\SetFigFont{12}{14.4}{rm}2}}}
\put( 97,605){\makebox(0,0)[lb]{\smash{\SetFigFont{12}{14.4}{rm}$s_1$}}}
\put(157,605){\makebox(0,0)[lb]{\smash{\SetFigFont{12}{14.4}{rm}$s_2$}}}
\put(227,605){\makebox(0,0)[lb]{\smash{\SetFigFont{12}{14.4}{rm}$s_3$}}}
\put(240,626){\makebox(0,0)[lb]{\smash{\SetFigFont{12}{14.4}{rm}1}}}
\end{picture}\end{array}  
\end{equation}
The Nahm data are $2\times 2$ in the left interval and $1\times 1$ in
the right interval. The boundary conditions imply that the $2\times 2$
data are nonsingular at the $s=s_2$ boundary with $T_i(s_2)_{_{2,2}}$
equal to the $1\times 1$ data.  The data have a pole at $s=s_1$. If
$s_2=s_3$ the data are reciprocal to an $\SU{3}$ $(2,[1])$ monopole solution.
The skyline is
\begin{equation}
\begin{array}{c}
\begin{picture}(205,85)(55,605)
\thinlines
\put( 60,620){\line( 1, 0){200}}
\put(100,620){\line( 0, 1){ 40}}
\put(100,660){\line( 1, 0){ 60}}
\put(160,660){\line( 0,-1){ 20}}
\put(160,640){\line( 1, 0){ 2}}
\put(162,640){\line( 0,-1){ 20}}
\multiput(160,640)(0.00000,-8.00000){3}{\line( 0,-1){  4.000}}
\put( 95,623){\vector(0,1){  36}}  
\put( 95,623){\vector(0,-1){  0}} 
\put(160,620){\line( 0,-1){  5}}
\put(100,620){\line( 0,-1){  5}}
\put(162,620){\line( 0,-1){  5}}
\put( 84,638){\makebox(0,0)[lb]{\smash{\SetFigFont{12}{14.4}{rm}2}}}
\put( 97,605){\makebox(0,0)[lb]{\smash{\SetFigFont{12}{14.4}{rm}$s_1$}}}
\put(142,605){\makebox(0,0)[lb]{\smash{\SetFigFont{12}{14.4}{rm}$s_2=s_3$}}}
\end{picture}\end{array}
\end{equation}
and the Nahm data are $2\times2$ matrices with a pole at $s=s_1$ but
not at $s=s_2$. This is the Nahm data for the $(2,[1])$ monopole  solution
originally due to Dancer \cite{D1,Dancer:kj}. It will be discussed in greater
detail in Sec. V.

\section{Parameters and Nahm data}
\label{parametersection}

\begin{figure}[t]
\begin{center}
\epsfig{file=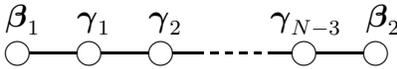}
\end{center}
\caption{SU(N) Dynkin diagram with labeled roots.\label{SUNCD}}
\end{figure}

We are interested in the case 
\begin{equation}
    \Phi_0 = {\rm diag}\, (s_R, s_0, s_0, \dots, s_0, s_L)
\end{equation}
where the middle $N-2$ eigenvalues of Eq.~(\ref{Phi0eigenvalues}) are
equal, thus breaking the gauge symmetry to
U(1)$\times$SU($N-2$)$\times$U(1).  There are two types of massive
monopoles and $N-3$ species of massless monopoles.  Our labeling of the
associated simple roots is indicated in the Dynkin diagram shown in
Fig.~\ref{SUNCD}.

With this symmetry breaking, the Nahm data consist of two triplets
$T_i^L(s)$ and $T_i^R(s)$ that are defined on the intervals
$(s_L,s_0]$ and $[s_0,s_R)$, respectively, as well as a set of complex
column vectors $a_p$ that form the jump data at $s_0$.

In this section, we will identify the parameters that we expect to
appear in our solutions and relate these to the Nahm data.  Our goal
is to study the $(2,[2], \dots, [2],2)$ solutions.  However, we will
find it useful to first examine the $(1,[1], \dots,[1],1)$ for SU($N$)
and the (2,[1]) for $\SU{3}$.  Both of these cases have cloud
parameters, and understanding these will be helpful 
in understanding the parameters of the $(2,[2], \dots, [2],2)$
solution.

The dimension of the moduli space is equal to four times the total
number of monopoles. However, there is one fewer moduli parameterizing
the $T_i$ and $a_p$ Nahm data.  This is because the modulus
parameterizing the overall global U(1) gauge rotation is associated
only with $T_0$ and plays no role in the gauge that we have chosen for
our Nahm construction.  The remaining global gauge freedom will be
present in our Nahm data, although for studying static nondyonic
solutions we will not need to explicitly display the corresponding
parameters and will usually work in a specific global gauge
orientation.

\subsection{$(1,[1], \dots,[1],1)$ solutions for SU($N$) }
\label{oneoneone}

These solutions are composed of two massive and $N-3$ massless
monopoles, and so form a $4(N-1)$-dimensional moduli space.  Six of
the parameters can be chosen to specify the positions of the massive
monopoles.  The remaining parameters include a number of global
gauge parameters as well as the parameters describing the
gauge-invariant properties of the cloud. 

Naively, one might expect the number of global gauge parameters to be
$(N-2)^2+1$, since this is the dimension of the unbroken
group. However, this cannot be correct in general; indeed, for large
$N$ this number far exceeds the total number of moduli available. In
the case of SU(4), it is easy to see that the full unbroken
U(1)$\times$SU(2)$\times$U(1) acts nontrivially on the solution, so
the naive formula is correct here and there are, indeed, five global gauge
parameters, leaving only a single cloud parameter.  

For $N>4$, the full set of solutions must include the subset obtained
by embedding the SU(4) solutions.  These embedded solutions will be
left invariant by a $\U{N-4}$ subgroup of the unbroken gauge group,
and thus will depend on
\begin{equation}
{\rm dim}\, [{\rm U}(1)\times {\rm SU}(N-2)\times{\rm
U}(1)/{\rm U}(N-4)] = 4N-11
\end{equation}
 global gauge parameters.  Together with the six position variables
and the cloud parameter that is inherited from the SU(4) solution,
this accounts for all $4(N-1)$ parameters.  Hence all the solutions
lie in the global group orbit of the embedded solutions. This
parameter-counting argument leaves open the possibility of a
disconnected $4(N-1)$-parameter family of solutions that are not SU(4)
embeddings, but the explicit Nahm construction of the fields rules
this out \cite{WY}.

With only one monopole of each type, the Nahm data are constants.
Their values on the left and right interval, ${\bf x}_L$ and ${\bf x}_R$
respectively, specify the positions of the two massive monopoles.  The
jump data at $s_0$ are a set of $N-2$ two-component $a_p$ that satisfy
\begin{equation}
     \delta {\bf T} = {\bf x}_L-{\bf x}_R = 
           \sum_{p=1}^{N-2} a_p^\dagger \sigmabf a_p  \, .
\label{su4jumpequation}
\end{equation}
Thus, they comprise $4(N-2) -3$ independent real numbers.  Some of
these are global gauge parameters associated with the U($N-2$) action
that takes $a_p$ to
\begin{equation}
      a'_p = U_{pq} a_q  \, .
\label{actionONa}
\end{equation}
Because there can only be two linearly independent two-component
$a_p$, they can, at most, span a two-dimensional subspace of the $(N-2)$-dimensional space on which the $U_{pq}$ act.  They are, therefore,
invariant under a U($N-4$) subgroup, so the number of gauge parameters
in the $a_p$ is ${\rm dim}\, [{\rm U}(N-2)/{\rm U}(N-4)] =
4N-12$.  This leaves one nongauge parameter in the jump data; this
parameter can be expressed as the single element of a $1\times 1$
``matrix'' $T_4$ defined by
\begin{equation}
     T_4 \id_2 + \delta T_i \sigma_i  
                =  \sum_{p=1}^{N-2} a_p \otimes a_p^\dagger \, .
\label{su4jump}
\end{equation}
The eigenvalues of the matrix on the right-hand side are obviously
positive.  This translates into the condition $p \equiv T_4 \ge R$, where $R
= |{\bf x}_L - {\bf x}_R|$.  It turns out that $p$ specifies the size
of the non-Abelian cloud.  When it takes its minimum value, $p=R$, the
right-side of Eq.~(\ref{su4jump}) has rank one.  There is then a
U($N-2$) transformation of the form of Eq.~(\ref{actionONa}) 
that leaves only a single nonvanishing $a_p$,
so the solution is, in fact, an embedding of a solution for SU(3) broken
to U(1)$\times$U(1).  Since there is no massless monopole in the SU(3)
theory, it is not surprising that this gives the minimal cloud.

It is instructive to compare these solutions with the $(1,1,\dots,1)$
solutions for maximally broken SU($N$).  In the latter case, there are
$N-1$ intervals, in each of which the Nahm data is constant with value
${\bf x}_n$.  At the boundary between two intervals, the jump data are
determined, up to an irrelevant phase, by the analogue of
Eq.~(\ref{su4jumpequation}).  This implies, in particular, that at the
$n$th boundary
\begin{equation}
     a_n^\dagger a_n =  |{\bf x}_n - {\bf x}_{n+1}|  \, .
\end{equation}

Examination of the spacetime fields for the case where the $|{\bf x}_n
- {\bf x}_m|$ are all large shows that the ${\bf x}_n$ are just the positions
of the massive monopoles.  In the limit where the middle $N-3$
monopoles become massless and the corresponding intervals in the Nahm
data shrink to zero width, the only remnant of these ${\bf x}_n$
is their effect on the jump data.  Thus, the effect of the U($N-2$)
action described above is to change the ``positions'' of the massless
monopoles, subject only to the constraint that
\begin{equation}
    p  =  \sum_{n=1}^{N-2} |{\bf x}_n - {\bf x}_{n+1}|.
\end{equation}
In other words, $p$ is a gauge invariant cloud parameter, but all the
other position moduli are acted on by the group action.

\subsection{$(2,[1])$ Dancer solutions for SU(3) }

The solutions for one massless and two massive monopoles in SU(3)
broken to SU(2)$\times$U(1) have been studied in detail 
\cite{D1,Dancer:kj,DL,I}.  Here we
choose our conventions so that the eigenvalues of the Higgs vacuum
expectation value are $(s_0, s_0, s_L)$, with $s_0 > s_L$.  The Nahm
data $T_i$ are Hermitian $2 \times 2$ matrices on the interval
$(s_L,s_0]$ that can be expanded as
\begin{equation}\label{DNansatz}
     T_i(s) = \frac{1}{2}{\bf C}_i(s)\cdot \btau  + R_i(s) \id_2 \, .
\end{equation}
Substituting this into the Nahm Eq.~(\ref{Ne}) show that the $R_i$
are independent of $s$ and the ${\bf C}_i$ satisfy
\begin{equation}\label{rotflow}
\frac{d}{ds}{\bf C}_i=
  \frac{1}{2}\epsilon_{ijk}{\bf C}_j\times {\bf C}_k \, .
\end{equation}
This means that the elements of the real symmetric matrix
\begin{equation}
    M_{ij} =  {\bf C}_i\cdot {\bf C}_j 
   - {1\over 3} \delta_{ij} {\bf C}_k \cdot {\bf C}_k
\end{equation}
are constants. If $A$ is the $s$-independent real orthogonal matrix that
diagonalizes $M$, the three vectors 
\begin{equation}
     {\bf B}_i(s) = A_{ji}{\bf C}_j(s) 
\end{equation}
are mutually orthogonal.  Furthermore, substitution into Eq.~(\ref{rotflow})
shows that the directions of these three vectors are independent of
$s$, so that we can write
\begin{equation}
     {\bf B}_i(s) = g_i(s) {\bf \hat e}_i
\end{equation}
(with no implied sum over $i$) where the ${\bf \hat e}_i$ are three
orthonormal vectors.  

The Nahm equation then reduces to the Euler-Poinsot equations
\begin{eqnarray}\label{EPeqn}
    {d g_1 \over ds} &=& g_2 g_3   \cr\cr
    {d g_2 \over ds} &=& g_3 g_1   \cr\cr
    {d g_3 \over ds} &=& g_1 g_2.  
\end{eqnarray}
These imply that the three quantities $\Delta_{ij} = g_i^2 -g_j^2$ are
constant, allowing us to order the ${\bf B}_i$ so that $|g_1(s)| \le
|g_2(s)| \le |g_3(s)|$.  Fixing one constant of integration by the
requirement that the $T_i$ have a pole at $s_L$ then gives
\begin{equation}
     g_i(s) = f_i(s-s_L; k, D)
\end{equation}
where the $f_i$ are Euler top
functions defined by 
\begin{eqnarray}
  f_1(u; k, D) &=& - D { {\rm cn}_k(Du) \over {\rm sn}_k(Du)} \cr\cr
  f_2(u; k, D) &=& \mp D { {\rm dn}_k(Du) \over {\rm sn}_k(Du)} \cr\cr
  f_3(u; k, D) &=& \mp {D \over {\rm sn}_k(Du) }
\label{topfunctions}
\end{eqnarray}
with the elliptic parameter $k$ lying in the range $0\le k \le 1$.  Note that 
there is a sign ambiguity in $f_2$ and $f_3$; to have a solution of the 
Euler-Poinsot equation, the same choice must be made for both functions.  
In our calculations we will have occasion to make use of both sign possibilities.

These top functions have poles at $u=0$ and at $u=2K(k)/D$, where
$K(k)$ is the complete elliptic integral of the first kind. We have
already arranged that the first pole is at the left boundary of the
Nahm interval, $s=s_L$. The second pole must lie beyond the right
boundary of the interval. This means that $2K/D>s_0-s_L$ and we define
a quantity $a$ by
\begin{equation}
    {1 \over a} = {4K(k) \over D} - 2(s_0 -s_L)   \, .
\label{defOFa}
\end{equation}
Roughly, $2a$ measures the size of the top functions at $s_0$.
Since there is a half in the expression for $T_i$ in
Eq.~(\ref{DNansatz}), $a$ determines the size of the matrix entries in
the data. We will see that it is a cloud parameter.

The final result for the Nahm data is 
\begin{equation}
     T_i(s) = \frac{1}{2}\sum_{jk} A_{ij} f_j(s-s_L; k, D) E_{jk}\tau_k 
           + R_i \id_2
\label{DancerNahmdata}
\end{equation}
where $E_{ij} \equiv (\hat e_i)_j$.  The expression depends on eleven
parameters.  The nature of some of these is evident.  The matrix
$E_{ij}$ depends on the three Euler angles that specify a global SU(2)
gauge transformations.  The three $R_i$ specify the position of the
center-of-mass.  To understand the remaining five, it is easiest to
first focus on the case where $A$ is a unit matrix.  From analysis of
asymptotic cases and examination of numerical solutions for the
spacetime fields \cite{Dancer:kj,DL,I} one finds that for large values
of $D$ there are two massive monopoles lying on the $z$-axis and
separated by a distance $D$.

The effect of $A$ is to rotate this configuration.  Hence, two of the
spatial Euler angles in $A$ are directly related to the positions of
the massive monopoles.  The third Euler angle, corresponding to
rotations about the axis joining the two monopoles, is a bit less
obvious.  Although one might expect a pair of monopoles to give an
axially symmetric configuration, we know from the SU(2) example that
this is not the case.  The asymmetry falls exponentially with the
monopole separation.  At infinite separation, where the axial symmetry
is recovered, rotations about this axis are equivalent to relative U(1)
global gauge transformations of the two monopoles.  For large, but
finite, separation it is still most useful to view the corresponding
Euler angle as being associated with a relative U(1) degree of
freedom.
 
We choose the final parameter to be the quantity $a$ that was defined
in Eq.~(\ref{defOFa}).  Examination of solutions shows that $a$
determines the size of the cloud.  For large $a$, the cloud is
approximately spherical with radius $a$.  As $a$ tends to infinity,
the second pole of the Nahm data approaches the boundary of the
interval, and the spacetime fields reduce to an embedding of the SU(2)
two-monopole solution.

It is useful to consider the corresponding $(2,1)$ solution for
$\SU{3}$ broken to $\U{1}\times\U{1}$.  Both species of monopoles are
massive, with the Dancer solution corresponding to the limit where the
second species becomes massless.  For the $(2,1)$ case, the Nahm data
in the left hand interval are precisely the Dancer Nahm data discussed
above.  The Nahm data in the right hand interval, which specify the
position of the monopole of the second type, are given by the vector
$((T_1)_{22},(T_2)_{22},(T_3)_{22})$.  One can show that when $a$ is
large it gives the approximate separation of the second type of
monopole from the center of mass of the two monopoles of the first
type. In the Dancer limit, this separation is the cloud parameter, but
the direction of this separation has no effect on the Dancer
solution. This example is discussed at length in \cite{HIM}.

\subsection{$(2,[2], \dots,[2],2)$ solutions for SU($N$) }

We now return to the case considered in Sec.~\ref{oneoneone}, but with
two monopoles, instead of one, of each type.  A
parameter-counting argument similar to that given in the previous
case shows that for large $N$ the generic solution is an
embedding of an SU(6) solution.  We will therefore start by focusing
on the case $N=6$.  The extension to larger $N$ is straightforward,
and we will discuss below the issues that arise when $N<6$. 

For the SU(6) case, there are four massive and ten massless monopoles,
and hence a 40-dimensional moduli space.  There are 17 global gauge
parameters, and we expect 12 more to specify the positions of the
massive monopoles.  This still leaves 11 parameters, enough to describe 
a much richer variety of solutions than were found in the previous cases.

By examining the Nahm data, we can clarify the interpretation of these
last 11 parameters.  On the left and right intervals the solutions of
the Nahm equations are just those found for the Dancer solution, but
with the arguments of the elliptic functions arranged so that the left
and right Nahm data have their poles at $s_L$ and $s_R$, respectively.
Thus,
\begin{eqnarray}
      T_i^L(s) &=& {1 \over 2} 
    \sum_{ij} A^L_{ij} f_j^L(s-s_L; k_L, D_L) \tau_j^L
            +  R_i^L \id_2      \cr
      T_i^R(s) &=& {1 \over 2}  
       \sum_{ij} A^R_{ij} f_j^R(s-s_R+2K/D_R; k_R, D_R) \tau_j^R
            +  R_i^R \id_2   \, .
\end{eqnarray}
where $f_j^L$ ($f_j^R$) are the top functions given in
Eq.~\ref{topfunctions} with the upper (lower) choice of sign; these
choices of signs will prove convenient in subsequent calculations. Here
we have defined two rotated triplets of Pauli matrices
\begin{eqnarray}
     \tau_i^L &=& E_{ij}^L \tau_j \cr 
     \tau_i^R &=& E_{ij}^R \tau_j
\end{eqnarray} 
where the matrices $E_{ij}^L$ and $E_{ij}^R$, each depending on three
Euler angles, encode the effect of two independent SU(2)
transformations on the standard set of Pauli matrices.

The $T_i^L(s)$ and the $T_i^R(s)$ each contain 11 parameters.  Six of
these specify massive monopole positions: $\bf R$, $D$, and two of the
Euler angles in the rotation matrix $A$. As discussed above, the third
Euler angle in $A$ is related to relative U(1) rotations of the
corresponding massive monopoles. In the Dancer $(2,[1])$ solution, a
cloud parameter $a$ was defined which depended on the distance
between the end of the interval and the pole. In the SU(6) context,
there are two such parameters, $a_L$ and $a_R$. These will be called
Dancer cloud parameters.

Finally, there are three SU(2) Euler angles in $E$.  In the Dancer
case, these were associated with global gauge transformations.  Now
that we have two sets of Dancer data, the relative SU(2) orientation
of the $\btau^L$ and $\btau^R$ is a physical quantity and is a
gauge-invariant property of the solutions. This is completely
analogous to the relative U(1) in the SU(2) two-monopole solutions.
On the other hand, a simultaneous SU(2) transformation of the
$\btau^L$ and $\btau^R$ is still equivalent to a global gauge
transformation of the solution.

The jump data consist of four $a_p$, each with four complex
components. As in Subsec.~\ref{ncogg}, we view these as being
two-component vectors whose components $a_{pj}$ are themselves
two-component spinors. There are 32 real parameters in the $a_p$. They
obey a jump equation 
\begin{equation}
     (\delta T_i)_{rs} \equiv (T_i^L)_{rs} -(T_i^R)_{rs} 
       = \sum_{p=1}^{4} a_{pr}^\dagger \sigma_i a_{ps} 
\label{su6jumpequation}
\end{equation}
which gives 12 real constraints. Furthermore, there is an U(4) action,
of the form of Eq.~(\ref{actionONa}), that gives rise to 16 of the
global gauge parameters. After subtracting these constraints and global
gauge parameters, we are left with four parameters.  These can be encoded
in a $2 \times 2$ matrix
\begin{equation}
     T_4 = p \id_2 + {\bf q} \cdot {\btau}
\end{equation}
obeying
\begin{equation}
     T_4\otimes \id_2 + \delta T_i \otimes \sigma_i  =  
               \sum_{p=1}^4  a_p \otimes a_p^\dagger  \, .
\label{su6jump}
\end{equation}
In the spinor notation
\begin{equation}
T_4=\sum_{p=1}^{4} a_{pr}^\dagger a_{ps}. 
\end{equation}

We will see that $p$ and $\bf q$ encode information about clouds that
are of a different type than the Dancer clouds; we will refer to these
as SU(4)-cloud parameters. Eq.~(\ref{su6jump}) forces $T_4\otimes \id_2
+ \delta T_i \otimes \sigma_i$ to have positive eigenvalues. This
constrains the cloud parameters. As we saw in Subsect.~\ref{oneoneone},
this happens in the $(1,[1], \dots,[1],1)$ case as well.
Equation~(\ref{su6jump}) is the same as Eq.~(\ref{su4jump}), except
that here we are dealing with matrices rather than numbers and, here,
the constraints do not, in general, have a simple expression in terms of
the other parameters.

There is some redundancy in the parameters that we have enumerated, in
that the effects of a common SU(2) transformation of the $\btau^L$ and
$\btau^R$ can be completely compensated by a U(4) action on the $a_p$,
which in turn can rotate the direction of $\bf q$.  Taking this factor
into account, we have the following 23 nongauge degrees of freedom:
\begin{enumerate}
\item[a)] 12 massive monopole position variables.
\item[b)] Two relative U(1) parameters, one for the $\bbeta_1$-monopoles
  and one for the $\bbeta_2$-monopoles.
\item[c)] Two Dancer cloud parameters, $a_L$ and $a_R$.
\item[d)] Two scalar SU(4) cloud parameters $p$ and $q=|{\bf q}|$.
\item[e)] Five parameters specifying the relative SU(2) orientations of
the triplets ${\btau}^L$ and ${\btau}^R$ and the vector $\bf q$.  
\end{enumerate}

When the gauge group is smaller than SU(6), the number of parameters
is reduced and additional constraints come into play.  After the
global gauge degrees of freedom are subtracted, the number of nongauge
parameters remaining is 14 for SU(3), 19 for SU(4), and 22 for SU(5).
This can be seen by counting the parameters arising from the
jump data.  For SU$(N)$, there are $(N-2)$ of the $a_p$, each
with eight real components, that are subject to a U($N-2$) gauge
action.  For $N=3$, 4, and 5, this gives seven, 12, and 15 nongauge
variables in the jump data.  However, there are 12 constraints imposed
by Eq.~(\ref{su6jumpequation}).  For SU(5), this means that $T_4$ has
only three free variables.  From Eq.~(\ref{su6jump}), whose right hand
side now has rank three, we see that these can be taken to be the
components of $\bf q$, with $p$ now fixed at its minimum value.  For
SU(4), counting arguments suggest that 
the constraints completely determine the jump data and that $p$
and $\bf q$ can be specified in terms of $T_i^L$ and $T_i^R$.  We will
see that the actual situation is more subtle, and that not all values
for $T_i^L$ and $T_i^R$ are possible; i.e., for some choices of Dancer
data the constraints cannot be solved. For
SU(3), there are more constraints than jump variables, so clearly
$T_i^L$ and $T_i^R$ cannot be independently chosen Dancer-type Nahm
data.

\section{$\lowercase{(k,[k], \dots,[k],k)}$ monopoles in SU($N$)}
\label{KKKsolutions}

In the previous section, we saw that, for the $(1,[1],
\dots,[1],1)$ and the $(2,[2], \dots,[2],2)$ cases, the solution of 
Nahm's equation can be reduced to an algebraic problem involving the
jump data and the previously known Nahm data for the SU(2) unit monopole
and the SU(3) $(2,[1])$-solution.  More generally, the Nahm data for
the $(k,[k], \dots,[k],k)$ solution in SU($N$) are clearly related to
the Nahm data for the $(k,[k-1],[k-2], \dots, [1])$ solution for
SU($k+1$) broken to U($k$).  We will extend the term Dancer solutions
to include these generalizations of the SU(3) case.

We will show, in this section, that something similar happens when
using the construction equation to calculate the spacetime fields from
the data.  Specifically, if the spacetime fields and the boundary
values of the Nahm data and of the construction equation solutions are
known for the Dancer problem, then the SU($N$) fields can be obtained
by purely algebraic means.  We will concentrate on the construction of
the Higgs field, but the generalization to the gauge potential $A_j$
is straightforward.

Throughout this section, we will assume that $N\ge 2k+2$.  The
solutions for $N < 2k+2$ can be obtained by constraining the Nahm data
so that the SU($2k+2)$ solution is equivalent to an embedding of a
solution from a smaller group.

The Nahm data on the left interval, $(s_L, s_0]$, are the $k\times k$
matrices $T^L_i(s)$ for the SU($k+1$) Dancer problem, with their
arguments chosen so that their poles are at $s_L$.  These define a
construction equation on this interval that has $k+1$ linearly
independent solutions $w^L_a(s; {\bf r})$, each of which is a
$2k$-component column vector.  These satisfy
\begin{equation}
          \int_{s_L}^{s_0} ds\, w_a^{L\dagger}(s; {\bf r}) w_b^L(s;
          {\bf r}) = \delta_{ab}
\end{equation}
and give rise to an SU($k+1$) Dancer solution with Higgs field
\begin{equation}
   \varphi^L_{ab}({\bf r}) = \int_{s_L}^{s_0} ds (s-s_0)
   w_a^{L\dagger}(s; {\bf r}) w_b^L(s; {\bf r} )  \, .
\end{equation}
Similarly, the Nahm data $T_i^R(s)$ on the right interval lead to
$k+1$ solutions $w^R_a$ that generate a Higgs field
\begin{equation}
   \varphi^R_{ab}({\bf r}) = \int_{s_0}^{s_R} ds (s-s_0)
   w_a^{R\dagger}(s; {\bf r}) w_b^R(s; {\bf r} ) \, .
\end{equation}

In addition, there are the $2k$-component $a_p$ ($p = 1, \dots, N-2$)
that comprise the jump data at $s_0$.  By exploiting the U($N-2$)
action of Eq.~(\ref{actionONa}), these can be chosen so that they satisfy a
relation of the form
\begin{equation}
      a_p^\dagger  a_q =  \lambda_p \delta_{pq}
\label{aorthonormality}
\end{equation}
for $p, q \le 2k$, while $a_p=0$ if $p> 2k$.  
Generalizing Eqs.~(\ref{su4jump}) and (\ref{su6jump}), we now define
\begin{equation}
     K = T_4 \otimes \id_2 + \delta T_i \otimes \sigma_i 
          = \sum_p a_p \otimes a_p^\dagger  \, .
\end{equation}
As we saw in Sec.~\ref{parametersection}, $T_4$ 
encodes the nongauge parameters in the $a_p$.  As long as all
$2k$ of the $\lambda_p$ are nonzero, $K$ is invertible, with 
\begin{equation}
   K^{-1} = \sum_{p=1}^{2k} {1 \over \lambda_p^2}\, 
     a_p \otimes a_p^\dagger  \, .
\label{deltaTinv}
\end{equation}

These data determine the form of the construction equation for the
SU($N$) theory.  We must find $N$ linearly independent solutions
$V_a$, each consisting of functions $v^L_a(s;{\bf r})$ and $v_a^R(s;{\bf
r})$ defined on the intervals $(s_L,s_0]$ and $[s_0,s_R)$,
respectively, and a set of $S_{ap}({\bf r})$ defined at the jump.
These must obey the orthonormality condition of Eq.~(\ref{genortho}).  We
proceed in two steps, first obtaining an intermediate set of solutions
that are linearly independent but not orthonormal, and then
orthonormalizing.

Thus, we define a set of $\tilde V_a(s;{\bf r})$ by requiring
\begin{eqnarray}
    \tilde v^L_a(s;{\bf r})  &=& \cases{ w_{a-(k+1)}^L(s;{\bf r}),  
                    &\quad $k+2 \le a \le  2k+2 $ \cr
                   0,    &  \quad  otherwise }  \cr\cr
    \tilde v^R_a(s;{\bf r})  &=& \cases{ -w_a^R(s;{\bf r})
                         ,    &\quad  $1 \le a \le k+1 $\cr
                    0,   &  \quad  otherwise.}
\end{eqnarray}
(Note that both $\tilde v^L_a$ and $\tilde v^R_a$ vanish if $a>2k+2$.)
The discontinuity conditions on the $\tilde V_a$ then take the form
\begin{equation}
    V^0_a({\bf r}) \equiv \tilde v^L_a(s_0;{\bf r}) 
   - \tilde v^R_a(s_0;{\bf r}) = \sum_p \tilde S_{ap}({\bf r}) a_p \, .
\end{equation}
The orthonormality condition on the $a_p$, Eq.~(\ref{aorthonormality}), 
determines the $\tilde S_{ap}$ for $p \le 2k$ to be
\begin{equation}
    \tilde S_{ap}  = 
       {1 \over \lambda_p}\, a^\dagger_p V^0_a  , \qquad p \le 2k  \, .
\end{equation}
The $\tilde S_{ap}$ for $p > 2k$ are undetermined; we make the choice
\begin{equation}
    \tilde S_{ap}  = \delta_{(a-2),p}  , \qquad p > 2k \, .
\end{equation}

These solutions are not properly orthonormalized.  Instead,
\begin{equation}
     \langle \tilde V_a | \tilde V_b \rangle  
          = \cases{ B_{ab}  &\quad $ a, b  \le 2k+2$ \cr
                   \delta_{ab} &\quad $ a, b > 2k+2$ \cr
                     0   &\quad  otherwise  }
\end{equation}
where $B$ is a $(2k+2) \times (2k+2)$ matrix with
\begin{eqnarray}
    B_{ab} &=& \delta_{ab} + \sum_{p=1}^{2k} {1 \over \lambda^2_p} \,
         \left( V^{0\dagger}_a a_p\right) \left( a^\dagger_p
               V^0_b \right) \cr
      &=& \delta_{ab} +V^{0\dagger}_a K^{-1}
           V^0_b  \, .
\label{defofB}
\end{eqnarray} 

$B$ is clearly a Hermitian matrix with positive eigenvalues, so
$B^{-1/2}$ exists.   We can therefore define a new set of
solutions by
\begin{equation}
     V_a  = \cases{ \tilde V_b (B^{-1/2})_{ba} , 
                  &\quad $ a \le 2k+2$ \cr
             \tilde V_a , &\quad $ a > 2k+2 \, .$ }
\end{equation}
It is easily verified that these $V_a$ are orthonormal.  Following
Eq.~(\ref{Higgs}), they give rise to a Higgs field
\begin{eqnarray}
    \Phi_{ab} &=& \langle V_a |s| V_b \rangle \cr
              &=& \langle  V_a |(s-s_0)| V_b \rangle 
                     + s_0 \,\delta_{ab}  \, .
\end{eqnarray}
In the second line, the first term gets contributions from the
$v^L_a(s; {\bf r})$ and the $v^R_a(s; {\bf r})$, but not from the
$S_{ap}$.  As a result, it can be nonzero only if $1 \le a, b \le
2k+2$.  Hence if $N > 2k+2$, the Higgs field is essentially an
embedding of an SU($2k+2$) field.  Similarly, one can show that
$A_i$ can be written as an embedded solution.  We lose little
generality, but gain simplification in the notation, by henceforth
assuming that $N=2k+2$.  This allows us to write
\begin{eqnarray}
     \Phi_{ab} &=& (B^{-1/2})_{ac} 
           \langle \tilde V_c |(s-s_0)| \tilde V_d \rangle
           (B^{-1/2})_{da} + s_0  \,\delta_{ab}  \cr\cr
      &=& (B^{-1/2})_{ac} \,\varphi_{cd} \, (B^{-1/2})_{da} 
            + s_0 \, \delta_{ab} 
\end{eqnarray}
where $\varphi$ is block diagonal:
\begin{equation}
\varphi=\left(\begin{array}{cc}\varphi^R&0\\0&\varphi^L\end{array}
    \right)\, .
\end{equation}

Because our main interest in this paper is in the massless monopole
clouds, we can simplify our analysis by restricting our attention to
the region of space lying outside the cores of the massive monopoles.
Outside these cores, there is a clear distinction between the massless
degrees of freedom associated with the unbroken gauge group and the
fields that acquire masses through the symmetry breaking.  When we
apply the Nahm construction to our problem, this distinction appears
as follows.  As we will show explicitly for the cases $k=1$ and $k=2$,
the solutions of the construction equation defined by the Dancer Nahm
data can be chosen so that one of the $v_a(s; {\bf r})$ is
concentrated near the side of the interval where the $T_i$ have a pole
and is exponentially small at the other side, while the remaining
$v_a(s; {\bf r})$ are all concentrated on the side away from the
pole.\footnote{This prescription for the $v_a(s; {\bf r})$ is a
choice of gauge.  It can be done locally without any problem, but
extending it over all of space introduces Dirac string
singularities.} The massive Higgs and gauge fields
involve integrals containing products of the first $v_a$ and one of
the latter, and so fall exponentially with distance from the nearest
massive monopole core.
Ignoring these exponentially small terms, we can write the
Dancer Higgs fields in the block diagonal forms
\begin{eqnarray}
    \varphi^R &=& \left(\matrix{ \phi^R  & 0   \cr
                                0  & \hat \varphi^R } \right)
                  +  \varphi^R_\infty   \cr\cr
    \varphi^L &=& \left(\matrix{ \hat\varphi^L  & 0   \cr
                                0  & \phi^L } \right)
                  +  \varphi^L_\infty   \, . 
\label{dancerblockphi}
\end{eqnarray}
Here $\varphi^R_\infty$ and $\varphi^L_\infty$ are diagonal matrices
corresponding to the Higgs expectation values of the Dancer solutions.
Because $\phi^R$ and $\phi^L$ are purely U(1) fields, it is easy
to see that they must be sums of poles of the form $\pm 1/r_n$ where
$r_n$ is the distance from the $n$th massive monopole and the upper
and lower sign apply to $\phi^L$ and $\phi^R$, respectively.  The
non-Abelian parts of the Dancer solutions are contained in the $k
\times k$ matrices $\hat \varphi^R$ and $\hat \varphi^L$.

A similar decomposition occurs for $B$, which can be written in the
block diagonal form
\begin{equation}
   B = \left(\matrix{ 1  & 0  & 0  \cr
                        0 &  \hat B  & 0 \cr
                        0 & 0 & 1 } \right) \, . 
\end{equation}
Here $\hat B$ is given by an expression of the same form as
Eq.~(\ref{defofB}), but with the indices only running over the $2k$
values corresponding to the $v_a$ that are nonvanishing near $s_0$. 
Finally, $\Phi$ can be written as 
\begin{equation}
   \Phi = \left(\matrix{ \phi^R  & 0  & 0  \cr
                        0 &  \hat \Phi  & 0 \cr
                        0 & 0 & \phi^L } \right)  
      + \Phi_\infty
\end{equation}
where the $2k \times 2k$ non-Abelian part of the Higgs field is
\begin{equation}\label{4.19}
     \hat \Phi = \hat B^{-1/2} \hat \varphi \hat B^{-1/2}
\label{hatPhiequation}
\end{equation}
and $\hat \varphi$ is a block diagonal matrix
\begin{equation}
\hat \varphi=\left(\begin{array}{cc} \hat\varphi^R
&0\\0&\hat\varphi^L\end{array}\right) \, .
\end{equation}

From Eq.~(\ref{hatPhiequation}) we can see quite clearly the role of
the cloud parameters in the $a_p$.  Note first that the scale of the
$V^0_a$ is set by the distances to the massive monopoles and by the
Dancer cloud parameters.  If the $a_p$ are all large compared to
these, the eigenvalues of $K^{-1}$ will be small, $\hat B$ will be
approximately a unit matrix, and the non-Abelian fields of the SU($N$)
solution will simply be those inherited from the two Dancer solutions.
If, instead, the $a_p$ are all small, the eigenvalues of $K^{-1}$ will
be large, $\hat B$ will be large, and the factors of $\hat B^{-1/2}$
will suppress the non-Abelian part of the SU($N$) fields.

These two cases correspond to being inside and outside a non-Abelian
cloud of a type similar to those found in the previously known
examples that were discussed in the introduction.  There are two new
features here, however.  First, there are now $k^2$ cloud parameters
contained in the $a_p$.  Second, there are additional cloud
parameters, of a somewhat different type, contained in the Dancer-type
solutions.  One of our goals in this work is to understand the
interplay between these different types of clouds.  In
Sec.~\ref{twotwotwosolutions}, we will examine these for the $k=2$
case, after first having obtained some some necessary results about
the SU(3) Dancer solution.  First, however, we will apply the
formalism that we have just developed to the $k=1$ case, verifying that we
recover the results of Ref.~~\cite{WY} for the $(1,[1],1)$ SU(4) case.

For $k=1$, the ``Dancer'' solution is simply the unit SU(2) monopole
solution.  We take the two massive monopoles to lie along the
$z$-axis, with the $\bbeta_1$-monopole at ${\bf x}_R=(0,0,-R/2)$ and
the $\bbeta_2$-monopole at ${\bf x}_L =(0,0,R/2)$.  From
Eq.~(\ref{su4jump}) we have $K = p + R \sigma_3$, and hence
\begin{equation}
    K^{-1} = { p - R \sigma_3 \over p^2 -R^2} \, .
\end{equation}

The solution of the construction equation for the unit monopole was
given in Sec.~\ref{Nahmsection}.  Using Eqs.~(\ref{VforUnitMono}) and
(\ref{normalizationfactor}), we can take the solutions of the
construction equations for the right and left Dancer problems to be
\begin{eqnarray}
     w^R_1 (s; {\bf r}) &=& \sqrt{2r_R} \left[1 - e^{-2r_R (s_R -s_0)}
        \right]^{-1/2} e^{-r_R (s_R -s)} \psi_R \cr
      w^R_2 (s; {\bf r})
        &=& \sqrt{2r_R} \left[1 - e^{-2r_R (s_R -s_0)} \right]^{-1/2}
        e^{-r_R (s -s_0)} \bar \psi_R \cr 
     w^L_1 (s; {\bf r}) &=&
        \sqrt{2r_L} \left[1 - e^{-2r_L (s_0 -s_L)} \right]^{-1/2}
        e^{-r_L (s_0-s)} \psi_L \cr 
      w^L_2 (s; {\bf r}) &=& \sqrt{2r_L}
        \left[1 - e^{-2r_L (s_0 -s_L)} \right]^{-1/2} e^{-r_L (s -s_L
          )} \bar \psi_L \, .
\label{oneoneoneWs}
\end{eqnarray}
Here ${\bf r}_R = {\bf r} - {\bf x}_R$ and ${\bf r}_L = {\bf r} - {\bf
x}_L$, while $\psi_R = \psi({\bf r}_R)$ and $\psi_L = \psi({\bf
r}_L)$ with $\psi({\bf r})$ given by Eq.~(\ref{spinordef}).

When $(s_R -s_0)r_R$ and $(s_0-s_L)r_L$ are both large, $w_1^R(s; {\bf
r})$ and $w_2^L(s; {\bf r})$ are concentrated near $s_R$ and $s_L$,
respectively, and are exponentially small at $s_0$, while $w_2^R(s;
{\bf r})$ and $w_1^L(s; {\bf r})$ peak at $s_0$.  Up to exponentially
small corrections, the corresponding Higgs fields take the block
diagonal form of Eq.~(\ref{dancerblockphi}), with $\phi^R = -1/2r_R$,
$\phi^L = 1/2r_L$, and $\hat\varphi^R$ and $\hat\varphi^L$ combining
to give
\begin{equation}\label{4.25}
     \hat \varphi = \left(\matrix{ 1/2r_R  & 0 \cr\cr
                               0 & - 1/2r_L } \right)  \, .
\end{equation}

To construct the matrix $\hat B$, we exclude the exponentially small
U(1) components of $V^0$, obtaining a reduced vector $\hat V^0$.
Using Eq.~(\ref{oneoneoneWs}), we have (up to exponentially small
corrections)
\begin{eqnarray}
    \hat V^0_1({\bf r}) &=& w_2^R(s; {\bf r}) 
                 = \sqrt{2r_R} \,\, \bar \psi_R  \cr\cr
     \hat V^0_2({\bf r}) &=& w_1^L(s; {\bf r}) 
                    = \sqrt{2r_L} \,\,  \psi_L  \, .
\end{eqnarray}
Hence, 
\begin{eqnarray}
   \hat B &=&  \id_2  + \hat V^{0\dagger} K^{-1} \hat V^0  \cr\cr
    &=&    \id_2  + {1\over p^2 -R^2}
      \left(\matrix{r_R \,\bar \psi_R^\dagger( p - R \sigma_3) \bar\psi_R 
   & 2\sqrt{r_L r_R}\, \bar \psi_R^\dagger( p - R \sigma_3) \psi_L \cr\cr
     2\sqrt{r_L r_R} \, \psi_L^\dagger(p - R \sigma_3) \bar \psi_R 
   & 2r_L\, \psi_L^\dagger(p - R \sigma_3) \psi_L } \right)  \, .
\label{oneoneoneB}
\end{eqnarray}
Using the form of the spinors given in Eq.~(\ref{spinordef}), 
together with the identities 
\begin{eqnarray}
     z_L + z_R &=& {r_R^2 -r_L^2 \over R}    \cr\cr
 z_L z_R &=& {1 \over 4 R^2} \left[ (r_R^2 -r_L^2)^2 - R^2 \right] \, ,
\end{eqnarray}
allows $\hat B$ to be simplified to
\begin{equation}
     \hat B = { r_L +r_R + p \over p^2 - R^2}
         \left[ p - R\,{\bf \hat q} \cdot \btau \right]
\label{Cresult}
\end{equation}
where the unit vector ${\bf \hat q}$ has components
\begin{equation}
    \hat q_i = \cases { \high{2r_i \over \sqrt{(r_L +r_R)^2 - R^2}} \, 
      & \quad    $ i = 1,2$  \cr\cr
          \high {r_L -r_R \over R} \, & \quad $i=3\, .$ } 
\end{equation}
(For $i=1$ and 2 we have used $x_L=x_R =x$ and $y_L=y_R=y$.)

Let $U$ be the unitary matrix that rotates $\tau_3$ to ${\bf \hat q}
\cdot \btau$, and define
\begin{equation}
    L = { p - R\tau_3 \over r_L +r_R + p}  \, .
\end{equation}   
Then $\hat B = U L^{-1} U^{-1}$ and 
\begin{equation}
     \hat \Phi = U L^{1/2} \left[
      \left({1 \over 4r_R} - {1 \over 4r_L} \right) \id_2 
        - \left({1 \over 4r_R} + {1 \over 4r_L} \right)
        {\bf \hat q} \cdot \btau \right]  L^{1/2} U^{-1}  \, .
\end{equation}
Taking into account that our cloud parameter $p$ is equal to the
quantity $2b +R$ of Ref.~\cite{WY}, we see that $\hat \Phi$ is the same, up
to a gauge transformation by $U$, as the previously obtained
expression.

\section{(2,[1]) monopole solutions and their fields}
\label{DancerSolutions}

In this section, we return to the $(2,[1])$ monopole solutions. The
Nahm data for these monopoles were discussed in
Sec.~\ref{parametersection}.  The Nahm construction of the monopole
fields corresponding to these data has not proven to be tractable.
However, as we describe in this section, it is possible to calculate
useful approximate fields in certain situations. Along with the
methods explained in the previous section, the $(2,[1])$ monopole
construction allows us to construct $(2,[2],\dots,2)$ monopole
solutions in $\SU{N}$. This will be considered in the next section.

Because we are primarily interested in the regions outside of the
massive monopole cores, we can work in the infinite mass limit with
data on the interval $(-\infty,s_0]$.  In order for the Euler top
functions introduced in Eq.~(\ref{topfunctions}) to be analytic on
this semi-infinite interval, the elliptic parameter $k$ must be unity,
implying\footnote{This $k=1$ limit is familiar as the
hyperbolic monopole data discussed by Dancer in Ref.~\cite{DL2}. There
is a subtle difference however, here we are interested in the infinite
mass limit and so it is the location $s=s_L$ pole that is being sent
to $s=-\infty$. In Ref.~\cite{DL2} the location of the second pole is
sent to $s=\infty$. This difference affects the signs of the
hyperbolic functions in the $k=1$ limit. Since it is convenient to
have all the signs identical in Eq.~\ref{htopfxns}, these signs have
been absorbed into the definition of $f_i^L$.}
\begin{eqnarray}\label{htopfxns}
f_1^L=f_2^L&=&-D\cosech{D(s-s_0-\epsilon)}\nonumber\\
    f_3^L&=&-D\coth{D(s-s_0-\epsilon)}
\end{eqnarray}
where $\epsilon = 1/2a$.

Further simplification occurs in two special cases.  The first, that
of minimal Dancer cloud, corresponds to $\epsilon \gg 1$.  The other,
that of large Dancer cloud, corresponds to $a/D \gg 1$.

\subsection{Minimal Dancer cloud}

If $\epsilon\rightarrow \infty$, the top functions become
$f_1^L=f_2^L=0$ and $f_3^L=D$ and the Nahm data of
Eq.~(\ref{DancerNahmdata}) reduce to
\begin{eqnarray}
     T_i(s) &=& \frac{1}{2}D_i \tau'_3  + R_i\, \id_2 \cr\cr
            &=& X^1_i \left({\tau'_3 + \id_2 \over 2} \right)
         + X^2_i \left({\tau'_3 - \id_2 \over 2} \right)
\end{eqnarray}
where the $\tau'_i = E_{ij}\tau_j$ are a set of rotated Pauli
matrices, $D_i = A_{i3} D$, and ${\bf X}^1 = {\bf R} + {\bf D}/2$ and
${\bf X}^2 = {\bf R} - {\bf D}/2$ are the positions of the two massive
monopoles.  

There is no pole in the Nahm data and the construction equation can be
put into a block diagonal form with a separate $2\times2$ block
corresponding to each of the two massive monopoles.  Its solutions can
be read off from the solutions to the one-monopole construction
equation found in Sec.~\ref{Nahmsection}.  The two solutions that are
nonzero at $s_0$ have boundary values
\begin{eqnarray}
    v_1^L(s_0) &=& \sqrt{2r_1}\, \psi({\bf r}_1)\otimes \chi_+  \cr\cr
    v_2^L(s_0) &=& \sqrt{2r_2}\, \psi({\bf r}_2)\otimes \chi_- 
\end{eqnarray}
where $r_1$ and $r_2$ are the distances to the massive monopoles and
$\chi_\pm$ are eigenvectors of $\tau'_3$ with eigenvalues $\pm 1$.  We
have inserted a superscript $L$ because the interval
$(-\infty,s_0]$ corresponds to the left
interval for the SU($N$) Nahm data.  After a constant factor is
extracted, the non-Abelian part of the Higgs field is
\begin{equation}
      \hat\phi^L=\left(\begin{array}{cc}
     {\high -\frac{1}{2r_1}}&{\high 0}\\{\high 0}&{\high
     -\frac{1}{2r_2}}\end{array}\right) \, . 
\end{equation}
This field corresponds to the massless monopole being coincident with
one of the two massive monopoles.

Following a similar procedure on the right interval, $[s_0, \infty)$,
leads to
\begin{eqnarray}
    v_1^R(s_0) &=& \sqrt{2r_1} \,\bar \psi({\bf r}_1)\, \chi_+  \cr\cr
    v_2^R(s_0) &=& \sqrt{2r_2} \,\bar \psi({\bf r}_2)\, \chi_-
\end{eqnarray}
and 
\begin{equation}
      \hat\phi^R=\left(\begin{array}{cc}
     {\high \frac{1}{2r_1}}&{\high 0}\\{\high 0}&{\high \frac{1}{2r_2}}\end{array}\right)  \, .
\end{equation}

\subsection{Small separation of the massive monopoles}

We now consider the case where the separation of the two massive
monopoles is small compared to the other scales of interest. 
In the limit $D\rightarrow 0$, the data are
spherically symmetric, with three equal top functions
\begin{equation}
   f_i^L=-\frac{1}{s-s_0-\epsilon}
\end{equation}
and Nahm matrices
\begin{equation}
   T_i(s) = -\frac{1}{2}\frac{1}{s-s_0-\epsilon} \tau'_i
\end{equation}
where now $\tau'_i = A^{ij}E^{jk} \tau_k$.  If we think of the
massless monopole as being positioned at
$((T_1)_{22},(T_2)_{22},(T_3)_{22})$, then it lies on a sphere of
radius $a = 1/2 \epsilon$ about the two massive monopoles and is
rotated by both the spatial rotations and the global gauge
transformations.  For simplicity, we begin by exploiting these
symmetries to position the massless monopole at $(0,0,a)$ (or
equivalently, to set $\btau'=\btau$), and calculate the fields along
the positive $z$-axis.

Hence, we want to solve
\begin{equation}
\frac{d}{ds}v = (r_i \id_2 - T_i ) \otimes \sigma_i v 
    \equiv (X-T) v  \, .
\end{equation}
In a basis where $\id_2\otimes \sigma_i$ is block diagonal with
diagonal blocks both equal to $\sigma_i$, 

\begin{equation}
X-T=\left(\begin{array}{cccc}\high\frac{1}{2(s-s_0-\epsilon)}+r
   &0&0&0\cr 0&\hskip -3pt\high-\frac{1}{2(s-s_0-\epsilon)}-r 
   &\hskip -3pt\high\frac{1}{s-s_0-\epsilon}&0\cr 0&\hskip -3pt\high\frac{1}{s-s_0-\epsilon}
    &\hskip -3pt\high-\frac{1}{2(s-s_0-\epsilon)}+r &0\cr 0&0&0&
    \hskip -4pt\high\frac{1}{2(s-s_0-\epsilon)}-r\end{array}\right)\/ .
\end{equation}

Two of the equations decouple, giving
\begin{equation}
   v_1= N_1 \sqrt{s_0-s + \epsilon} \, e^{r(s-s_0-\epsilon)}
\left(\begin{array}{c}1\\0\\0\\0\end{array}\right)
\end{equation}
and
\begin{equation}
   v_4=N_4 \sqrt{s_0-s+\epsilon}\, e^{-r(s-s_0-\epsilon)}
\left(\begin{array}{c}0\\0\\0\\1\end{array}\right).
\end{equation}

Since $v_1$ is a decaying solution as $s\rightarrow -\infty$, it is an
acceptable solution to the construction solution.  Choosing $N_1$ so
that $\int_{-\infty}^{s_0}v_1^tv_1 \,ds =1$, we have
\begin{equation}
    v_1(s_0)  =  {\sqrt{2}\,r \over \sqrt{r+a} }
   \left(\begin{array}{c}1\\0\\0\\0\end{array}\right) \, .
\label{leftv1boundary}
\end{equation}

The corresponding component of the Higgs field is
\begin{eqnarray}
\phi_{11}&=&\int_{-\infty}^{s_0}s\,v_1^\dagger v_1\,ds \nonumber\\
&=&s_0-\frac{1}{r}+\frac{1}{2(r+a)}  \, .
\label{Dancerphi11}
\end{eqnarray}

The other two equations are coupled. We substitute
\begin{equation}
v_2(s) =\left(\begin{array}{c}0\\p(s)\\q(s)\\0\end{array}\right)
\end{equation}
and find
\begin{eqnarray}
2(s-s_0-\epsilon)(\dot{p}+rp)+p-2q&=&0\nonumber\\
2(s-s_0-\epsilon)(\dot{q}-rq)+q-2p&=&0
\end{eqnarray}
with overdots denoting differentiation with respect to $s$. These give
the second order equation
\begin{equation}
    \ddot{p}+\frac{2}{s-s_0-\epsilon}\dot{p}-r^2p
  +\frac{r}{s-s_0-\epsilon}p-\frac{3}{4(s-s_0-\epsilon)^2}p=0
\end{equation}
which is a Bessel equation. We are interested in the decaying solution
\begin{equation}
   p= - N_2 \frac{1}{[r(s_0-s+\epsilon)]^{3/2}}e^{r(s-s_0-\epsilon)}
\end{equation}
with corresponding $q$:
\begin{equation}
    q= N_2 \left(\frac{2}{\sqrt{r(s_0-s+\epsilon)}}
  +\frac{1}{[r(s_0-s +\epsilon)]^{3/2}}\right)e^{r(s-s_0-\epsilon)}\, .
\end{equation}

The normalization constant is fixed by requiring
\begin{equation}
  1  = \int_{-\infty}^{s_0} (p^2+q^2)ds 
      = N_2 (2I_3 +4I_2 +4 I_1)
\end{equation}
where 
\begin{equation}
I_n=\int^{s_0}_{-\infty}{\frac{1}{[r(s_0-s +\epsilon)]^n}
    e^{2r(s-s_0-\epsilon)}ds} \, .
\end{equation}
Integrating by parts shows that
\begin{equation}
I_{n+1}=\frac{1}{n}\left(\frac{1}{r^{n+1}\epsilon^n}
  e^{-2r\epsilon}-2I_n\right)
\end{equation}
for $n>0$.  $I_0$ can be integrated exactly and all the $I_1$ terms
cancel, leading to 
\begin{equation}
   v_2(s_0) = {\sqrt{2}\,r \over \sqrt{r+a}}
   \left(\begin{array}{c}0\\0\\1\\0\end{array}\right)
    + {\sqrt{2}\,a \over \sqrt{r+a}}
   \left(\begin{array}{c}0\\-1\\1\\0\end{array}\right) \, .
\label{leftv2boundary}
\end{equation}
The corresponding component of the Higgs field is
\begin{equation}
  \phi_{22}= \int_{-\infty}^{s_0} (p^2+q^2)s\, ds
   =  s_0-\frac{1}{2(a+r)} \, .
\label{Dancerphi22}
\end{equation}

Equations~(\ref{Dancerphi11}) and (\ref{Dancerphi22}) give the
diagonal elements of the Higgs field.  The off-diagonal elements
clearly vanish, since $v_1(s)$ and $v_2(s)$ are pointwise orthogonal.
Hence, we have found that the non-Abelian part of the Higgs field is
\begin{equation}
  \hat\phi^L=\left(\begin{array}{cc}{\high s_0-\frac{1}{r}
 +\frac{1}{2(a+r)}}&{\high 0}\\{\high 0}
   &{\high s_0-\frac{1}{2(a+r)}}\end{array}\right)
\label{leftDancerHiggs}
\end{equation}
when $r$ and $a$ are both much bigger than both the monopole core
size and the monopole separation. This expression was previously derived
in \cite{I} using symmetry arguments. It exhibits the role played by the
cloud. Inside the cloud, $r\ll a$ and
\begin{equation}
  \hat\phi^L\approx\left(\begin{array}{cc}{\high
  s_0+\epsilon-\frac{1}{r}}&{\high 0}\\ 
  {\high 0}&{\high s_0-\epsilon}\end{array}\right)
\label{insideDancerHiggs}
\end{equation}
which is the field of the two massive monopoles. The only effect of
the massless monopole is to modify the monopole mass. However, for
$r\gg a$
\begin{equation}
 \hat\phi^L\approx
   \left(\begin{array}{cc}{\high s_0-\frac{1}{2r}}&{\high 0}\\
   {\high 0}&{\high s_0-\frac{1}{2r}}\end{array}\right) \, .
\label{outsideDancerHiggs1}
\end{equation}
Thus, at large distance, there is a ${\rm diag}\,(1/2r,-1/2r)$
contribution to the field from the massless monopole.  In other words,
the massless monopole charge screens the massive monopole charges at a
distance scale of roughly $a$.

Up to now we have restricted ourselves to the case where $\bf r$ is
along the positive $z$-axis and the SU(2) orientation of the Dancer
cloud is such that $\btau'=\btau$.  The more general case can be
obtained by applying appropriate symmetry transformations to our
solution.  Suppose that $\btau' = U \btau U^{-1}$, where 
\begin{equation}\label{5.28}
    U = \left(\matrix {f  & g \cr -g^* & f^* } \right) \, .
\end{equation} 
For an arbitrary position, not necessarily on the $z$-axis, 
Eqs.~(\ref{leftv1boundary}) and (\ref{leftv2boundary}) are then replaced by 
\begin{eqnarray}
    v_1(s_0) &=&  {\sqrt{2}\,r \over \sqrt{r+a} }
   \left(\begin{array}{c} \lambda_+ \psi({\bf r}) \\
       \lambda_- \psi({\bf r})\end{array}\right)  \cr 
   v_2(s_0) &=& {\sqrt{2}\,r \over \sqrt{r+a}}
   \left(\begin{array}{c}-\lambda_-^* \psi({\bf r}) \\
       \lambda_+^* \psi({\bf r}\end{array}\right)
    + {\sqrt{2}\,a \over \sqrt{r+a}}
   \left(\begin{array}{c}g\\-f\\f^*\\g^*\end{array}\right)
\end{eqnarray}     
where the two-component vector $\psi({\bf r})$ is defined by
Eq.~(\ref{spinordef}) and 
\begin{eqnarray}
    \lambda_+ &=&  \sqrt{r-z \over 2r}
           \left[ {x-iy \over r-z}\, f + g\right] \cr\cr
    \lambda_- &=&  \sqrt{r-z \over 2r}
            \left[f^* - {x-iy \over r-z}\, g^* \right]  \, .
\end{eqnarray}
Equation~(\ref{leftDancerHiggs}) for the Higgs field remains unchanged.  

Well outside the cloud, where $r \gg a$, the second term in
$v_2^L(s_0)$ is suppressed by a factor of $a/r$.  If this term is 
ignored, then a linear combination of the $v_j^L(s)$ gives an
alternative basis, $v_j^{'L}(s)$, for which the boundary values
\begin{eqnarray}
   v_1^{'L}(s_0) &\approx& \sqrt{2r}
   \left(\begin{array}{c} \psi({\bf r}) \\ 0 \end{array}\right) \cr\cr\cr
   v_2^{'L}(s_0) &\approx& \sqrt{2r}
   \left(\begin{array}{c} 0 \\ \psi({\bf r}) \end{array}\right) 
\end{eqnarray}
are independent of the SU(2) rotation to leading order in $a/r$.  The
leading approximation to the Higgs field,
Eq.~(\ref{outsideDancerHiggs1}), is unaffected by this change of basis
and so is independent of the SU(2) orientation parameters in $U$.

Proceeding in the same manner on the interval $[s_0, \infty)$ leads
to 
\begin{eqnarray}
    v_1(s_0) &=&  {\sqrt{2}\,r \over \sqrt{r+a} }
   \left(\begin{array}{c} \mu_+ \bar\psi({\bf r}) \\
       \mu_- \bar\psi({\bf r})\end{array}\right)  \cr 
   v_2(s_0) &=& {\sqrt{2}\,r \over \sqrt{r+a}}
   \left(\begin{array}{c}-\mu_-^* \bar\psi({\bf r}) \\
       \mu_+^* \bar\psi({\bf r}\end{array}\right)
    + {\sqrt{2}\,a \over \sqrt{r+a}}
   \left(\begin{array}{c}g\\-f\\f^*\\g^*\end{array}\right)
\end{eqnarray} 
with
\begin{eqnarray}
    \mu_+ &=&  \sqrt{r-z \over 2r}
           \left[ f + {x+iy \over r-z}\, g \right] \cr
    \mu_- &=&  -\sqrt{r-z \over 2r}
            \left[ {x+iy \over r-z}\, f^* + g^*\right] \, .
\end{eqnarray}
The corresponding Higgs field has a non-Abelian component
\begin{equation}
  \hat\phi^R=\left(\begin{array}{cc}{\high s_0 +\frac{1}{r}
 -\frac{1}{2(a+r)}}&{\high 0}\\{\high 0}
 &{\high s_0+\frac{1}{2(a+r)}}\end{array}\right)  \, .
\label{rightDancerHiggs}
\end{equation}

\section{Explicit $(2,[2],[2],[2],2)$ solutions in SU(6)}
\label{twotwotwosolutions}

In this section, we will use the results of Sec.~\ref{DancerSolutions} on
SU(3) Dancer solutions, together with the general formalism developed
in Sec.~\ref{KKKsolutions},  to obtain $(2, [2], [2], [2], 2)$ solutions for
SU(6) broken to U(1)$\times$SU(4)$\times$U(1).  As previously, we will
concentrate on the region outside the massive monopole cores.  Hence,
our primary interest will be in the non-Abelian part $\hat \Phi$ of
the Higgs field.  From Eq.~(\ref{4.19}), this is given by
\begin{equation}
    \hat \Phi = \hat B^{-1/2} \hat\varphi \hat B^{-1/2} = \hat B^{-1/2}
      \left( \matrix{ \hat\varphi^R & 0 \cr 0 & \hat\varphi^L} \right)
       \hat B^{-1/2}
\end{equation}
where the $2 \times 2$ matrices $\hat\varphi^L$ and $\hat\varphi^R$
are obtained from the Dancer solutions corresponding to the Nahm data
on the left and right intervals $(s_1,s_0]$ and $[s_0,s_2)$.  The $4
\times 4$ matrix $\hat B$ is
\begin{equation} 
    \hat B = {\rm I}_4 +\hat V^{0\dagger}(s_0) K^{-1}\hat V^0(s_0) \, .
\label{hatBforSec6}
\end{equation} 
Here $\hat V^0(s_0)$ is obtained from the solutions of construction
equations for the right and left Dancer solutions, while
\begin{equation}
    K = [p {\rm I}_4 + {\bf q}\cdot \btau]  \otimes {\rm I}_2
       + [T_i^L(s_0) - T_i^R(s_0)] \otimes \sigma_i  \, .
\end{equation} 
For the remainder of this section we will omit the argument of
$\hat V^0$; it should always be understood to be $s_0$.
Note that $\hat V^0$ depends on $\bf r$, while $K$ does
not.  

In principle, as long as the left and right Dancer solutions are of
one of the two types studied in Sec.~\ref{DancerSolutions}, we can
obtain expressions for $\hat \Phi$ that are exact, up to exponentially
small corrections, outside the massive cores.  However, our primary
goal is to understand the role played by the cloud parameters.  In
particular, we want to see how the clouds affect the non-Abelian
magnetic fields.   This can be summarized by a position-dependent
magnetic charge $Q_{NA}$.  For a spherically symmetric configuration,
this can be immediately read off from the non-Abelian part of the Higgs
field, with 
\begin{equation}
    \hat \Phi = -{Q_{NA} \over 2r} + O\left({1 \over r^2}\right) \, . 
\end{equation}
For less symmetric configurations, where the magnetic field has
contributions from Coulomb fields centered at several different points
${\bf x}_j$, $Q_{NA}$ is obtained by adding the coefficients of the
$1/|{\bf r} - {\bf x}_j|$ terms in the Higgs field.  With this goal in
mind, we focus on several limiting cases in which the existence of
small parameters makes it particularly easy to pick out the $1/r$
terms in the Higgs field.

\subsection{Two minimal Dancer clouds}

We first consider the case where both the left and right data correspond
to Dancer clouds of minimal size.  The massive monopoles are located
at 
\begin{eqnarray}
     {\bf x}_1 &=& {\bf R}^R +{\bf D}^R    \cr
     {\bf x}_2 &=& {\bf R}^R -{\bf D}^R    \cr
     {\bf x}_3 &=& {\bf R}^L + {\bf D}^L   \cr
     {\bf x}_4 &=& {\bf R}^L -{\bf D}^L    \, .
\end{eqnarray}
We fix the SU(2) orientation of the SU(4) cloud so that ${\bf q}=
(0,0,q)$, but for the moment leave those of the two Dancer clouds
arbitrary.  This leads to 
\begin{equation}
   K = (p {\rm I}_2+ q \tau_3) \otimes {\rm I}_2  
       + (R_i + D^L_i\tau_3^L - D^R_i \tau_3^R) \otimes \sigma_i  
\label{minimalcloudT}
\end{equation}
where ${\bf R} = {\bf R}^L -{\bf R}^R$.

The solutions of the construction equation yield the boundary values
\begin{equation}
   \hat V^0_j = \sqrt{2r_j} \,s_j \otimes \chi_j  \, .
\end{equation}
Here, the quantities $s_j$ and $\chi_j$ are defined by 
\begin{eqnarray}
   \hat V^0_1(s_0) &=&\sqrt{2r_1} \, \bar\psi({\bf r}_1) \otimes
         \chi^R_+  \cr  
   \hat V^0_2(s_0) &=& \sqrt{2r_2} \, \bar\psi({\bf r}_2) \otimes 
         \chi^R_-  \cr
   \hat V^0_3(s_0) &=&\sqrt{2r_3} \, \psi({\bf r}_3) \otimes \chi^L_+  \cr 
   \hat V^0_4(s_0) &=&\sqrt{2r_4} \, \psi({\bf r}_4) \otimes \chi^L_-  \ 
\label{minimalVs}
\end{eqnarray}
where $\psi({\bf r})$ is given by Eq.~(\ref{spinordef}) and $\chi_\pm^L$ 
($\chi_\pm^R$) are eigenvectors of $\tau_3^L$ ($\tau_3^R$) with
eigenvalues $\pm 1$.  These solutions also lead to 
\begin{equation}
    \hat \varphi = {1 \over 2r} \, {\rm diag}\, (1,1,-1,-1) \, .
\end{equation}

We will analyze in detail three special cases:

1) SU(2) orientations of both Dancer clouds and the SU(4) cloud all
   aligned. 

2) ${\bf x}_1 ={\bf x}_3$ and ${\bf x}_2 ={\bf x}_4$, with the Dancer
clouds having identical SU(2) orientations.

3) Large SU(4) clouds: $p\pm q \gg |{\bf x}_i -{\bf x}_j|$ for all $i$,
   $j$. 

\subsubsection{All cloud SU(2)'s aligned}
\label{alignedMinimal}

\begin{figure}[t]
\begin{center}
\epsfig{file=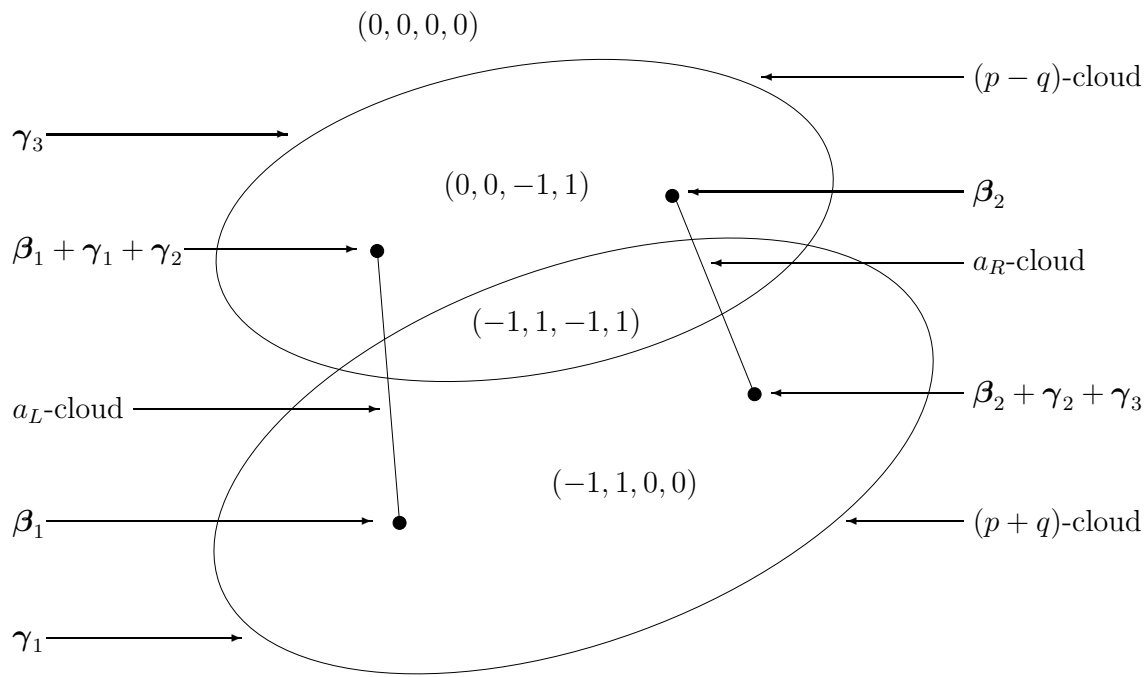,width=15cm}
\vskip 1cm
\caption{
A solution with two minimal Dancer clouds and all cloud SU(2)
orientations aligned.  The clouds are labelled both by the relevant
distance scale and by the associated massless monopole.  As described
in Sec.~\ref{candm}, the location of the massless monopoles at the end
of the Dancer clouds is a gauge-dependent choice.  The diagonal
elements of $Q_{NA}$, which is assumed to be a diagonal matrix, are
shown for each of the regions defined by the clouds.\label{minoverFig}}
\end{center}
\end{figure}

If the SU(2) orientations of the Dancer and SU(4) clouds are all
identical, so that $\tau_3^R = \tau_3^L = \tau_3$, then 
\begin{equation}
     K = \left({{\rm I}_2 + \tau_3 \over 2}\right) \otimes
      \left[ (p+q)\id_2 + ({\bf x}_3 - {\bf x}_1) \cdot \bsigma \right]
       + \left({{\rm I}_2 - \tau_3 \over 2}\right) \otimes
    \left[ (p-q)\id_2 + ({\bf x}_4 - {\bf x}_2) \cdot \bsigma \right] \, .
\end{equation} 
Because the $\hat V^0_j$ are all eigenvectors of $\tau^3$, the problem
reduces to two independent problems, each equivalent to the SU(4) $(1, [1],
1)$ problem discussed in Sec.~\ref{KKKsolutions}.  One yields an
ellipsoidal SU(4) cloud with cloud parameter $p+q$ that encloses
monopoles 1 and 3, while the other gives a cloud with parameter $p-q$
enclosing monopoles 2 and 4.  Because the non-Abelian fields of the
two clouds lie in mutually commuting SU(2) subgroups of the unbroken
SU(4), the only interactions between the clouds are those from the
short range interactions involving the massive monopole cores.  Hence,
the two clouds can be disjoint, one enclosed within the other, or, as
shown in Fig.~\ref{minoverFig}, overlapping.

\subsubsection{Coincident massive monopoles}
\label{coincidentMinimal}

If there are pairs of coincident massive monopoles, with ${\bf x}_1
={\bf x}_3$ and ${\bf x}_2 ={\bf x}_4$, then ${\rm D}^L = {\rm D}^R
={\rm D}$ and ${\bf R}=0$.
If we also choose $\tau_3^L$ and $\tau_3^R$
to be identical, although not necessarily equal to $\tau^3$, then
\begin{equation}
     K = (p {\rm I}_2+ q \tau_3) \otimes {\rm I}_2 \, . 
\label{coincidentT}
\end{equation} 
Substituting this, together with Eq.~(\ref{minimalVs}), into
Eq.~(\ref{hatBforSec6}) yields
\begin{equation}
    \hat B = {\rm I}_2 + {1 \over p-q} M^- + {1 \over p+q} M^+
\label{coincidentB}
\end{equation}
where
\begin{equation}
   M^\pm = 2\sqrt{r_i r_j} \,s_i^\dagger s_j \,
      (\chi_i^\dagger P_\pm \chi_j)
\label{coincidentmatrix}
\end{equation}
and $P_{\pm} = ({\rm I}_2 \pm \tau_3)/2$.

The $s_j$ are fixed by the massive monopole positions and our definitions
of $\psi({\bf r})$ and $\bar\psi({\bf r})$.  We fix the $\chi_j$ by
writing
\begin{equation}
     \chi^L_+ =\chi^R_+ = \left(\matrix{c \cr d}\right)
    \qquad \qquad
     \chi^L_- =\chi^R_- = \left(\matrix{-d \cr c}\right)
\label{explicitChi}
\end{equation}
with $c$ and $d$ real.\footnote{For simplicity of notation, we have
chosen the phases of the $\chi_j$ so that only real quantities appear
in Eq.~(\ref{explicitChi}).  One can verify that this has no effect on
our final results.}  To normalize the spinors, we require $c^2 +d^2
=1$.  It then follows that
\begin{equation}
     c^2 - d^2 = {1 \over 2} {\rm Tr} \, \tau_3 \tau^L_3 
          \equiv \cos \alpha \, .
\label{cosAlphaDef}
\end{equation}

Substitution of the explicit expressions for the $s_j$ and the $\chi_j$
into Eq.~(\ref{coincidentmatrix}) then gives
\begin{equation}
     M^- = \left( \matrix{ u^2  & uv f^* & 0 & uv g^* \cr\cr
             uv f & v^2 & -uv g^* & 0 \cr\cr
             0 & -uv g & u^2 & uv f \cr\cr
            uv g & 0 & uv f^* & v^2 } \right)
\end{equation}
where 
\begin{eqnarray} 
     u &=& d \sqrt{r_1}  \cr
     v &=& c \sqrt{r_2}  \cr
     f &=& \psi({\bf r}_1)^\dagger \psi({\bf r}_2)  \cr
     g &=& \psi({\bf r}_1)^\dagger \bar \psi({\bf r}_2) \, .
\end{eqnarray}
(Note that $|f|^2 +|g|^2 = 1$.)  The expression for $M^+$ can be
obtained by making the substitutions $c \rightarrow d$ and $d
\rightarrow -c$.  The eigenvalues of $M^-$ are doubly degenerate, with
two being zero and two equal to $r_1 d^2 + r_2 c^2$; those of $M^+$ are zero
and $r_1 c^2 + r_2 d^2$.

Let us now define 
\begin{eqnarray}
    \lambda_-  &=&  {r_1 d^2 + r_2 c^2 \over p-q} 
       = {(r_1 + r_2) + (r_2-r_1) \cos\alpha \over 2(p-q)}    \cr\cr
    \lambda_+  &=&  {r_1 c^2 + r_2 d^2 \over p+q} 
       = {(r_1 + r_2) - (r_2-r_1) \cos\alpha \over 2(p+q)}   \, .
\label{lambdaPMdef}
\end{eqnarray}
We can then write
\begin{eqnarray}
    \hat B &=& {\rm I}_4 
       + \lambda_- \left({2M^- \over  {\rm Tr}\, M^-}\right) 
       + \lambda_+ \left({2M^+ \over  {\rm Tr}\, M^+}\right)  \cr\cr
      &=& {\rm I}_4 + \lambda_- \Pi^- + \lambda_+ \Pi^+
\end{eqnarray}
where $\Pi^\pm$ are both projection operators.  The
spaces onto which they project are not in general mutually orthogonal;
their overlap is measured by
\begin{equation} 
       {\rm Tr} \, \Pi^+ \Pi^-
       = {2 (r_1 -r_2)^2 \sin^2\alpha \over 
         4 r_1 r_2 + (r_1 -r_2)^2 \sin^2\alpha} \le 2 \, .
\end{equation}
This is small when $r_1$ and $r_2$ are both large compared to the
monopole separation $D$, so far from the massive monopoles $\Pi^+
\approx {\rm I}_4 - \Pi^-$ and the eigenvalues of $\hat B$ are
approximately $1 + \lambda_-$ and $1 + \lambda_+$, both being doubly
degenerate.
For simplicity, let us assume that $ p +q \gg D$.  In this case,
$\lambda_+$ is non-negligible only in the large distance region where
$\Pi^+ \approx {\rm I}_4- \Pi^-$.  Hence, $\hat B$ can be approximated
(everywhere) by
\begin{equation} 
  \hat B = (1 + \lambda_-) \Pi^- + (1 + \lambda_+)({\rm I}_4- \Pi^-) \, . 
\end{equation}

The degree to which the non-Abelian charges are shielded, depends on
the magnitudes of $\lambda_-$ and $\lambda_+$.  If both are much less
than unity, $\hat B$ is approximately a unit matrix and there is no
shielding by the SU(4) cloud.  The field corresponds to an effective
non-Abelian magnetic charge 
\begin{equation}
     Q_{\rm NA} = {\rm diag}\, (-1,-1,1,1) \, .
\label{unshieldedminimal}
\end{equation}
If both are much greater than unity, there is complete shielding of
the non-Abelian charge and $Q_{\rm NA}=0$. In the intermediate case,
where $\lambda_- \gg 1$ but $\lambda_+ \ll 1$,
\begin{equation}
  \hat \Phi \approx ({\rm I}_4- \Pi^-) \hat \varphi ({\rm I}_4- \Pi^-) \, .
\end{equation}
This corresponds to a  charge that has only two
nonzero eigenvalues, with the other two vanishing as a result of
shielding by the cloud:
\begin{equation}
     Q_{\rm NA} = {\rm diag}\, (-1,0,1,0) \, .
\label{partialNAcharge}
\end{equation}

The boundaries between the regions corresponding to these three
possibilities are roughly given by the surfaces $\lambda_+=1$ and
$\lambda_-=1$.  Given our assumption that $ p +q \gg D$, the former
surface is approximately spherical, with radius $p+q$.  The topology
of the latter surface depends on the magnitude of $p-q$:

1) If $p-q > D(1 +|\cos\alpha|)/2$, the surface $\lambda_-=1$ encloses
all of the massive monopoles.  

2) If $D(1 -|\cos\alpha|)/2<p-q < D(1 +|\cos\alpha|)/2$, the surface
$\lambda_-=1$ encloses only one of ${\bf x}_1$ and ${\bf x}_2$.

3) If $p-q < D(1 -|\cos\alpha|)/2$, then $\lambda_-$ is always greater
than unity.  In this case, the unshielded charge of
Eq.~(\ref{unshieldedminimal}) never occurs.

\begin{figure}[t]
\begin{center}
\epsfig{file=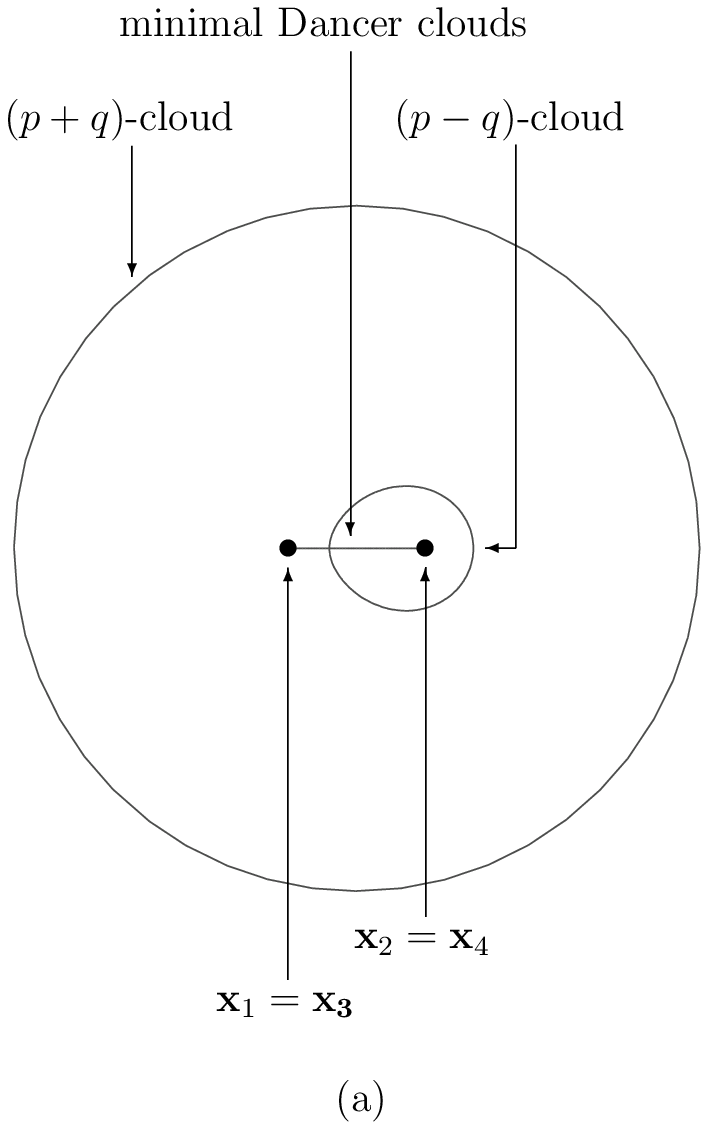,width=6.5cm}\quad\qquad\epsfig{file=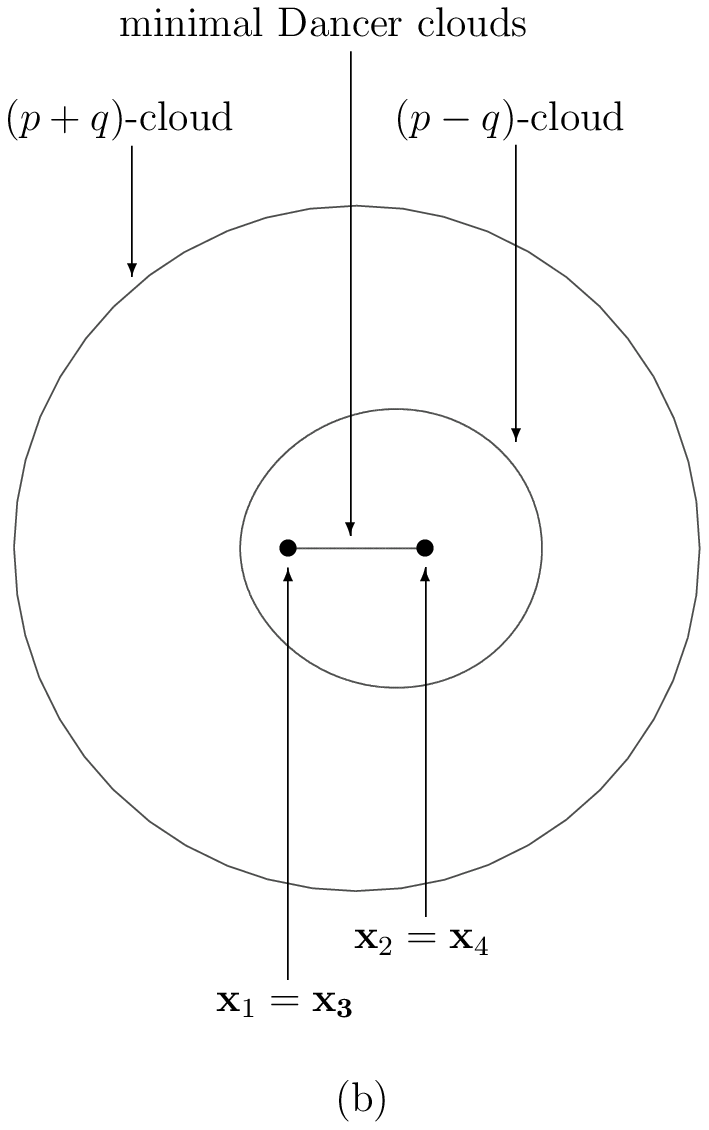,width=6.5cm}
\vskip .75cm
\caption{Schematic illustration of a solution with minimal Dancer
clouds and coincident massive monopoles, as discussed in
Subsec.~\ref{coincidentMinimal}. In both figures the $(p+q)$-cloud is
roughly defined by the curve $\lambda_+=1$ with $p+q=5D$ and the
$(p-q)$-cloud by the curve $\lambda_-=1$, where in (a) $p-q=1.2D$ and in
(b) $p+q=2.2D$. In each case $\alpha=4\pi/3$ and there are coincident
$\betabf_1$- and $\betabf_2$-monopoles at ${\bf x}_1={\bf x}_3$ and
at ${\bf x}_2={\bf x}_4$. The massless monopole locations are not
shown, but calculating $Q_{NA}$ in different regions indicates that
there are $\gammabf_1$- and $\gammabf_3$-monopoles on the minimal
Dancer clouds, a $(\bgamma_2+\bgamma_3)$-monopole on the $(p-q)$-cloud and
a $(\bgamma_1+\bgamma_2)$-monopole on the $(p+q)$-cloud.\label{notalignedFig}}
\end{center}
\end{figure}

Possibilities 1) and 2) are illustrated in Fig.~\ref{notalignedFig}.

\subsubsection{Large SU(4) clouds}

If $p +q$ and $p-q$ are both much larger than all of the $|{\bf x}_i -
{\bf x}_j|$, then the first term on the right hand side of
Eq.~(\ref{minimalcloudT}) dominates and $K$ is well approximated by
Eq.~(\ref{coincidentT}) and $\hat B$ by Eq.~(\ref{coincidentB}).  At
short distances (all $r_j \ll p \pm q$), $\hat B$ is approximately a
unit matrix and one sees the unshielded charge of
Eq.~(\ref{unshieldedminimal}).  At much larger distances, the $r_j$
are all approximately equal.  Equation~(\ref{coincidentmatrix}) then
gives
\begin{equation}
     M^- \approx 2r \left( \matrix{ c_R^2  & c_R\, d_R  & 0 & 0 \cr\cr
             c_R\, d_R  & d_R^2 & 0 & 0 \cr\cr
             0 & 0 & c_L^2 & c_L \,d_L \cr \cr
            0 & 0 & c_L \,d_L  & d_L^2 } \right)
\end{equation}
and
\begin{equation}
     {M^+\over 2r}  \approx {\rm I}_4 - {M^- \over 2r} \, .
\end{equation}
(Here we have used the notation of Eq.~(\ref{explicitChi}), but have
allowed for the possibility that the SU(2) orientations of the left
and right Dancer clouds might be different.)  It follows that 
\begin{equation}
     \hat B \approx {p -q +2r \over p-q} \, {M^- \over 2r}
   + {p +q +2r \over p+q} \,\left({\rm I}_4- {M^- \over 2r}\right) \, .
\end{equation}
Because $M^-/2r$ is a projection operator, $\hat B^{-1/2}$ is easily
calculated, and one obtains
\begin{equation}
   \hat \Phi \approx {1 \over 2r} \,
       U  \left( \matrix{ \high{p-q \over p-q +2r}  & 0&0&0 \cr\cr
          0&\high{p+q \over p+q +2r}&0&0 \cr\cr 
          0&0&-\high{p-q \over p-q +2r}&0 \cr\cr
          0&0&0&-\high{p+q \over p+q +2r} } \right) U^{-1}
\end{equation} 
where $U$ is a unitary matrix that diagonalizes $M^-$.  Hence, the
non-Abelian magnetic charge is completely shielded for $r \gg p+q$,
partially shielded, as in Eq.~(\ref{partialNAcharge}), for $p+q \gg r
\gg p-q$, and unshielded for $r \ll p-q$.

\subsection{Two large Dancer clouds}

We now consider the case where both Dancer clouds are large.  For
convenience, we choose our spatial axes so that the ``left cloud'',
corresponding to the Nahm data ${\bf T}^L(s)$, is centered at
$(0,0,-R/2)$ and the ``right cloud'', obtained from ${\bf T}^R(s)$ is
centered at $(0,0,R/2)$.  We denote the distances from these two
centers by $r_L$ and $r_R$.  The cloud parameters for the two clouds
are $a_L$ and $a_R$, while the SU(2) orientations for the two clouds
are encoded in the rotated triplets of Pauli matrices $\btau^L$ and
$\btau^R$.

The matrix $K$ is then 
\begin{equation}
   K = [p \id_2 + {\bf q}\cdot \btau]  \otimes \id_2
          + R \id_2\otimes \sigma_3 
           + {a_R}\, \tau^R_i \otimes \sigma_i
           + {a_L}\, \tau^L_i \otimes \sigma_i  \, .
\label{KforTwoLargeD}
\end{equation}
The SU(4) cloud parameters $p$ and $\bf q$ must be such that the
eigenvalues of $K$ are all positive.  We will write $q = |{\bf
q}|$. 

Three special cases are particularly easy to analyze:

1) Large SU(4) cloud: $p\pm q \gg R,\, a_L,\, a_R$.

2) Widely separated Dancer clouds: $R \gg a_L,\, a_R$.

3) Two concentric large Dancer clouds: $a_R \gg a_L$, with $R=0$.

\subsubsection{Large SU(4) cloud}
\label{largecloudsubsec}

The eigenvalues of $K$ are 
\begin{eqnarray}
     \lambda_1 &=& p + q + O(R, a_L, a_R)  \cr 
     \lambda_2 &=& p + q + O(R, a_L, a_R)  \cr
     \lambda_3 &=& p - q + O(R, a_L, a_R)  \cr
     \lambda_4 &=& p - q + O(R, a_L, a_R)  \, .
\end{eqnarray}
Our assumption that $p\pm q \gg R,\, a_L,\, a_R$ ensures that these
eigenvalues are all positive.

If $r \ll p-q$, the various $\hat V^0_a$ are of order $\sqrt{r_R}$,
$\sqrt{r_L}$, $\sqrt{a_R}$, or $\sqrt{a_L}$, depending on whether $\bf
r$ is outside or inside a Dancer cloud.  Because these are all small
compared to the eigenvalues of $K$, the second term in
Eq.~(\ref{hatBforSec6}) can be neglected, $\hat B \approx \id_4$
and $\hat\Phi$ is a composite of two Dancer fields.  If instead $r
\gg p-q$, then $\bf r$ is well outside both Dancer clouds and the
$\hat V^0_a$ are all insensitive to the SU(2) orientation of the
Dancer clouds.  We can, therefore, orient the SU(4) cloud parameters so
that ${\bf q} = (0,0,q)$ (thus making $K$ diagonal) and at the
same time choose
\begin{eqnarray}
     \hat V^0_1  &\approx& \sqrt{2r}
   \left(\begin{array}{c} \bar\psi({\bf r}) \\ 0 \end{array}\right)
   \qquad\qquad
      \hat V^0_2  \approx \sqrt{2r}\left(\begin{array}{c} 0 
        \\ \bar\psi({\bf r})
   \end{array}\right) \cr\cr\cr 
     \hat V^0_3 &\approx& \sqrt{2r}
   \left(\begin{array}{c} \psi({\bf r}) \\ 0 \end{array}\right)
   \qquad\qquad
      \hat V^0_4  \approx\sqrt{2r} \left(\begin{array}{c}
      0 \\ \psi({\bf r}) 
       \end{array}\right) \, . 
\end{eqnarray}
(We have used the fact that $r_L \approx r_R$ in this region.)  To
leading approximation, this makes $\hat B$ diagonal.  Since $\hat
\phi$ is also diagonal outside the Dancer clouds,
\begin{equation}
   \hat \Phi \approx {1 \over 2r} 
         \left( \matrix{ \high{p-q \over p-q +2r}  & 0&0&0 \cr\cr
          0&\high{p+q \over p+q +2r}&0&0 \cr\cr 
          0&0&-\high{p-q \over p-q +2r}&0 \cr\cr
          0&0&0&-\high{p+q \over p+q +2r} } \right)  \, .
\end{equation} 

Thus, there are effectively two SU(4) clouds, one at $r \approx p+q$ 
and one at $r \approx p-q$.  In the region outside both clouds, there
are no non-Abelian Coulomb magnetic fields.  In the intermediate
region between the two clouds, the field corresponds to a non-Abelian
magnetic charge 
\begin{equation}
    Q_{NA} = \rm{diag}\, (0,-1,0,1)  \, .
\label{intermedQ}
\end{equation} 
In the region inside the inner SU(4) cloud, but still outside the two
Dancer clouds, the fields correspond to 
\begin{equation}
    Q_{NA} = \rm{diag}\,(-1,-1,1,1)  \, .
\label{interiorQ}
\end{equation} 

\subsubsection{Widely separated Dancer clouds}
\label{wideSepSubsec}

If $R \gg a_L, a_R$, the eigenvalues of $K$ are 
\begin{eqnarray}
     \lambda_1 &=& p + q +R   +O(a_L,a_R)  \cr 
     \lambda_2 &=& p + q -R   +O(a_L,a_R) \cr
     \lambda_3 &=& p - q +R   +O(a_L,a_R) \cr
     \lambda_4 &=& p - q -R   +O(a_L,a_R)   \, .
\label{widelySepEvalues}
\end{eqnarray}
In order that these all be positive, $p \pm q \ge R$ [up to
$O(a_L,a_R)$ corrections], which implies that $\lambda_1 ,\, \lambda_3 \ge
2R$. 

If the SU(2) basis is chosen so that ${\bf q} =
(0,0,q)$, then, up to $O(a_L,a_R)$ corrections,  $K$ is diagonal with 
\begin{equation}
    K^{-1} = {\rm diag}\, \left( {1 \over \lambda_1}, 
    {1 \over \lambda_2},  {1 \over \lambda_3},  {1 \over \lambda_4} 
     \right)  \, .
\end{equation}
In the region outside both Dancer clouds ($r_R \gg a_R$, $r_L \gg
a_L$), the SU(2) orientation of the Dancer clouds is
irrelevant and we can take
\begin{eqnarray}
     \hat V^0_1  &\approx& \sqrt{2r_R}
   \left(\begin{array}{c} \bar\psi({\bf r}_R) \\ 0 \end{array}\right)
   \qquad\qquad
      \hat V^0_2  \approx \sqrt{2r_R}\left(\begin{array}{c} 0 
        \\ \bar\psi({\bf r}_R)
   \end{array}\right) \cr\cr\cr 
     \hat V^0_3 &\approx& \sqrt{2r_L}
   \left(\begin{array}{c} \psi({\bf r}_L) \\ 0 \end{array}\right)
   \qquad\qquad
      \hat V^0_4  \approx\sqrt{2r_L} \left(\begin{array}{c}
      0 \\ \psi({\bf r}_L) 
       \end{array}\right) \, .
\end{eqnarray}

Examining the factors that enter into $\hat B$, we see that it is
composed of two interlocking $2 \times 2$ blocks.  One (containing the
11-, 13-, 31-, and 33-elements) is of the form of
Eq.~(\ref{oneoneoneB}) that we encountered in the construction of the
SU(4) $(1, [1], 1)$ solution, except that the cloud parameter is now
$p+q$.  The other (lying in the second and fourth rows and columns) is
similar, except with cloud parameter $p-q$.

Now consider the region inside one of the Dancer clouds.  Choosing the
right cloud, for definiteness, we have ${\bf r}_L \approx (0, 0, R)$.
The two $\hat V^0_a$ from the right-hand construction problem are
\begin{equation}
     \hat V^0_1 = O(r_R/\sqrt{a_R})
   \qquad\qquad
      \hat V^0_2 \approx \sqrt{2a_R}   \left(\begin{array}{c} 
    g \\ -f  \\  f^*\\  g^*  \end{array}\right)
\end{equation}
where, as in Eq.~(\ref{5.28}), $f$ and $g$ are elements of the
SU(2) matrix that relates the $\btau^R$ to the standard set of
$\btau$.  Since this region is well outside the left Dancer cloud, the
SU(2) orientation of that cloud is irrelevant and the remaining $\hat
V^0_a$ can be taken to be
\begin{eqnarray}
     \hat V^0_3 &\approx& \sqrt{2R}
   \left(\begin{array}{c} \psi({\bf r}_L) \\ 0 \end{array}\right)
       \approx \sqrt{2R}
         \left(\begin{array}{c}1 \\  O(a_L/R,a_R/R)\\ 0\\  0 \end{array}\right)
       \cr\cr
      \hat V^0_4  &\approx&\sqrt{2R} \left(\begin{array}{c}
      0 \\ \psi({\bf r}_L)  \end{array}\right) 
      \approx \sqrt{2R}
      \left(\begin{array}{c}0 \\  0\\ 1\\ O(a_L/R,a_R/R) \end{array}\right) \, .
\end{eqnarray}

We now use the facts that $\lambda_1^{-1}$ and $\lambda_3^{-1}$ are at
most $O(1/R)$ and that $\lambda_2^{-1}$ and $\lambda_4^{-1}$ can be at
most $O(1/a_R)$ before our approximations break down.  Together with
the above expressions for the $\hat V^0_a$, these imply that all the
elements of $\hat B_{ab}$ are of order unity or smaller. In fact, the
off-diagonal elements are $O(\sqrt{a_R/R})$, except for $\hat
B_{12}=(\hat B_{21})^* = O(r_R/a_R)$.  Hence, there is no significant
modification of the Dancer Higgs fields inside the Dancer cloud.

\begin{figure}[t]
\begin{center}
\epsfig{file=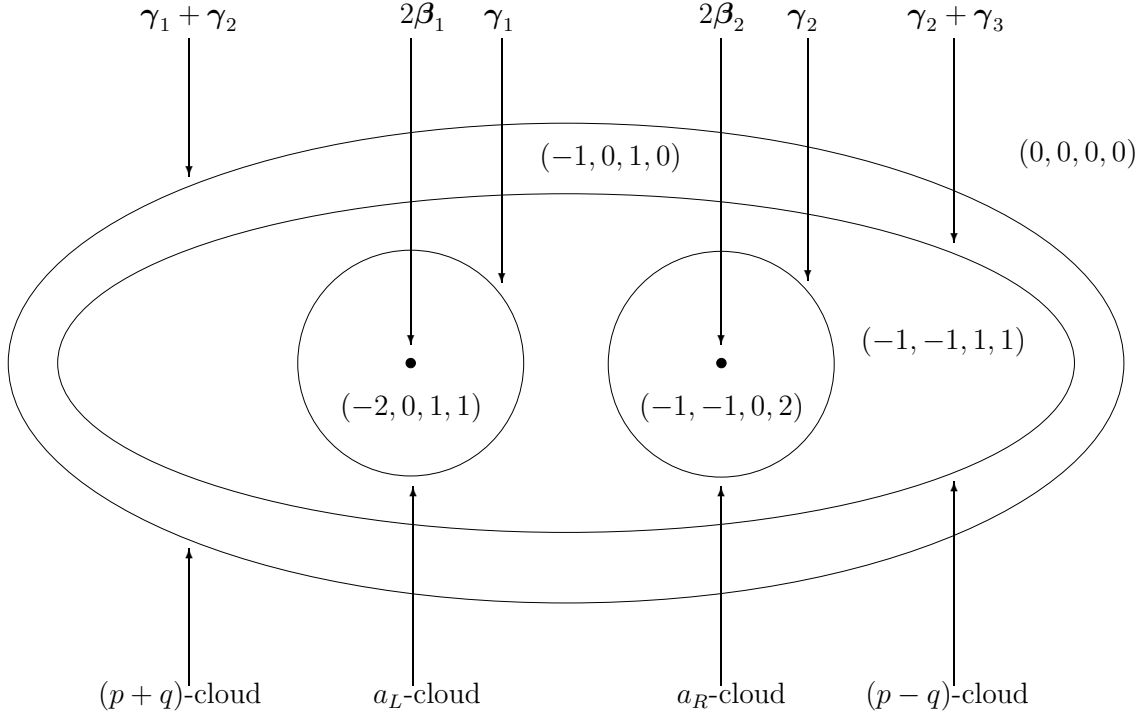,width=15cm}
\vskip 1cm
\caption{Schematic illustration of a solution with two widely-separated large
Dancer clouds. The clouds are labelled both by the relevant distance
scale and by the associated massless monopole.  The diagonal elements
of $Q_{NA}$, which is assumed to be a diagonal matrix, are shown for
each of the regions defined by the clouds.\label{largeRFig}}
\end{center}
\end{figure}

The results of this analysis are summarized in Fig.~\ref{largeRFig},
where we indicate the regions delineated by the various clouds and
show the value of $Q_{NA}$ in each.

\subsubsection{Two concentric large Dancer clouds}
\label{concentricSubsec}

We now consider two concentric Dancer clouds, with $a_R\equiv a \gg
a_L$.  Without loss of generality, we can choose the SU(2) orientation
of the right Dancer cloud so that the $\btau^R$ are the
standard Pauli $\btau$.  The orientation of the SU(4) cloud, encoded
in $\bf q$, and of the left Dancer cloud are arbitrary; 
the orientation of the latter will
actually play no role in our considerations.  Finally, because these 
solutions are spherically symmetric, it is sufficient to examine the
fields along the positive $z$-axis.

The matrix $K$ is 
\begin{equation}
    K \approx [p \id_2 + {\bf q}\cdot \btau]  \otimes \id_2
         + {a_R}\, \tau^R_i \otimes \sigma_i  \, .
\end{equation}
To leading order, its eigenvalues (although not its eigenvectors) are
independent of the direction of $\bf q$.  They are
\begin{eqnarray}
     \lambda_1 &=& p + q + {a}  +O(a_L)  \cr \cr
     \lambda_2 &=& p - {a} + \sqrt{4a^2 + q^2}   +O(a_L) \cr\cr
     \lambda_3 &=& p - q + {a}+O(a_L) \cr\cr
     \lambda_4 &=& p - {a} - \sqrt{4a^2 + q^2}   +O(a_L) \, . 
\end{eqnarray}
These obey $\lambda_1 \ge \lambda_2 \ge \lambda_3 > \lambda_4$, with
the first two relations being equalities only if $q=0$.  Note that the
last relation can never be an equality, so only $\lambda_4$ can
vanish.  Furthermore, once $p$ is chosen to make $\lambda_4$ positive,
the remaining $\lambda_j$ are all $O(a)$ or larger.

We will examine separately the region well outside the Dancer cloud,
$r \gg a$, and the region well inside the cloud, $r \ll a$.  Outside the 
cloud, it is possible to choose a basis so that the
$\hat V^0_j$ along the $z$-axis are 
\begin{eqnarray}
    \hat V^0_1 &=& \sqrt{2r}\, (0,1,0,0)^t  \cr\cr
    \hat V^0_2 &=& \sqrt{2r}\, (0,0,0,1)^t  \cr\cr
    \hat V^0_3 &=& \sqrt{2r}\, (1,0,0,0)^t  \cr\cr
    \hat V^0_4 &=& \sqrt{2r}\, (0,0,1,0)^t  \, .
\end{eqnarray}
In this basis, 
\begin{equation}
    \hat \varphi = {1\over 2r}  {\rm diag}\, (1,1,-1,-1) \, .
\label{outsideDancerHiggs2}
\end{equation}

There several possibilities to consider, depending on the magnitudes
of the $\lambda_j$.  If these eigenvalues are all $O(a)$ or smaller,
then in this region all of the eigenvalues of $\hat B$ will be
$O(r/a)$ and $\hat \Phi$ will be $O(a/r^2)$, corresponding to complete
shielding of the non-Abelian magnetic charge.  If, instead, the
$\lambda_j$ are all much greater than $a$, the ``large SU(4) cloud''
analysis given above applies.  There are effectively two SU(4) clouds,
one of radius $p+q$ and other of radius $p-q$.  The effective
non-Abelian magnetic charge vanishes outside both, is given by
Eq.~(\ref{intermedQ}) between the two, and is given by
Eq.~(\ref{interiorQ}) for $p-q \gg r \gg a$.

The only remaining possibility is that $\lambda_1 \approx \lambda_2
\approx p+q \gg a$ while $\lambda_3$ and $\lambda_4$ are $O(a)$ or
smaller.  As before, the non-Abelian charge is completely shielded for
$r \gg p+q$.  In calculating the fields at shorter distance to
leading order in $a/r$, we can approximate $K^{-1}$ by $\Pi
K^{-1} \Pi$, where
\begin{equation}
    \Pi \approx { \id_2+ \hat {\bf q} \cdot \btau \over 2} \otimes \id_2
\end{equation}
projects onto the subspace spanned by the eigenvectors corresponding
to $\lambda_3$ and $\lambda_4$.  It is easy to see that by a change of
basis that mixes $\hat V^0_1$ with $\hat V^0_2$ and $\hat V^0_3$ with
$\hat V^0_4$ one can obtain new vectors such that $\hat V^{'0}_1$ and
$\hat V^{'0}_3$ lie in the subspace onto which $\Pi$ projects while
$\hat V^{'0}_2$ and $\hat V^{'0}_4$ lie in the orthogonal subspace.
(This change of basis leaves Eq.~(\ref{outsideDancerHiggs2})
unchanged.)  In this basis $\hat B$ is approximately the identity in
the 2-4 subspace, but has two large eigenvalues (of order $r/a$) in
the 1-3 subspace.  As a result, two of the eigenvalues of the Higgs
field are shielded, so that the only large components are in the 2-4
subspace and the effective magnetic charge is given by
Eq.~(\ref{intermedQ}).

We now turn to the region well inside the Dancer cloud, $r \ll a$,
although still with $r \gg a_L$.  To leading order in $r/a$ the $\hat
V^0_j$ along the positive $z$-axis can be taken to be
\begin{eqnarray}
    \hat V^0_1 &=& \sqrt{2r}\, (0,0,0,1)^t  \cr\cr
    \hat V^0_2 &=& \sqrt{2a}\, (0,-1,1,0)^t  \cr\cr
    \hat V^0_3 &=& \sqrt{2r}\, (1,0,0,0)^t  \cr\cr
    \hat V^0_4 &=& \sqrt{2r}\, (0,0,1,0)^t  
\end{eqnarray}
while to the same order
\begin{equation}
    \hat \varphi = {1\over 2r}  {\rm diag}\, (2,0,-1,-1) \, .
\label{insideoutsideDancerHiggs}
\end{equation}

If all of the $\lambda_j$ are $O(a)$ or larger, then $\hat B \approx
I$ in this region and there is no shielding of non-Abelian magnetic
charge.  The only other possibility is that $\lambda_4 \equiv \Delta
\ll a$, with the remaining $\lambda_j$ being at least $O(a)$.  We now
examine this second case.  Let $u$ be the eigenvector of $K$
with eigenvalue $\lambda_4$, and define $w_j = u^\dagger \hat V^0_j$.
Using the above expressions for the $\hat V^0_j$, we have
\begin{equation}
    \hat B_{ij} = \delta_{ij} + {1 \over \Delta} w_i^\dagger w_j 
       + c \, \delta_{i2}\delta_{j2} + O(\sqrt{r/a})
\end{equation}
where $c$ is of order unity.  Next, note that $w_2$ is larger than 
the other $w_j$ by a factor of order $\sqrt{r/a}$ that arises from the
relative magnitudes of the $\hat V^0_j$.  Hence, to leading order the
term containing $c$ can be included in the $w_i^\dagger w_j$ term, giving
\begin{equation}
    \hat B_{ij} = \delta_{ij} + {(1+ c\, \Delta) \over \Delta} 
      w_i^\dagger w_j  +  O(\sqrt{r/a})  \, .
\end{equation}
Inverting this matrix gives
\begin{eqnarray}
    (\hat B^{-1})_{ij} &=& \delta_{ij} - {w_i^\dagger w_j \over |w|^2}
       + { \Delta \,w_i^\dagger w_j \over (1+ c \,\Delta)|w|^4} 
       +  O(\sqrt{r/a})  \cr \cr
    &=& \delta_{ij} - \delta_{i2}\delta_{j2} + O(\sqrt{r/a})
\end{eqnarray}
where $|w|^2 = w_i^\dagger w_i$.  Recalling the vanishing of
$\hat\varphi_{22}$ in Eq.~(\ref{insideoutsideDancerHiggs}), we see
that to leading order in $r/a$ there is no modification of the Higgs
field.

It is straightforward to extend this analysis to the region, $r \ll
a_L$, inside the smaller cloud and to show that to leading order the
SU(4) clouds do not modify the Higgs field there.  This result does not
depend on the relative SU(2) orientation of the two Dancer clouds.

\begin{figure}[t]
\begin{center}
\epsfig{file=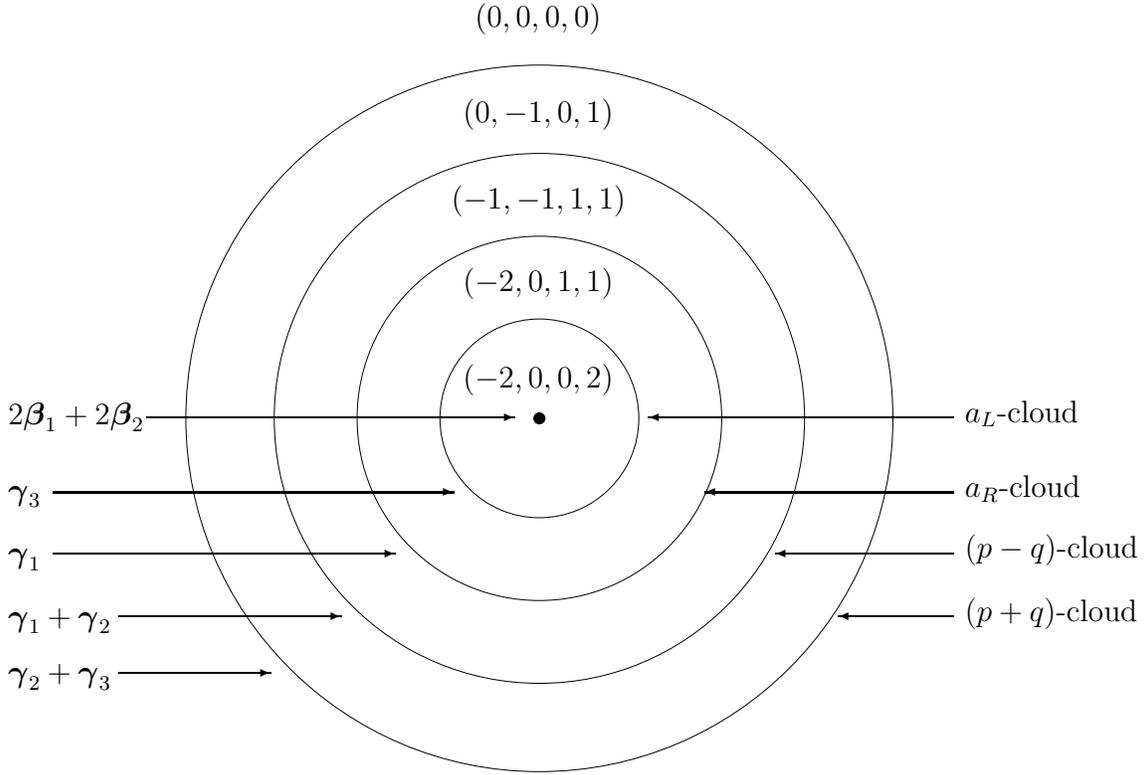,width=15cm}
\vskip 1cm
\caption{
Schematic illustration of a solution with two concentric large
Dancer clouds. The clouds are labelled both by the relevant distance
scale and by the associated massless monopole.  The diagonal elements
of $Q_{NA}$, which is assumed to be a diagonal matrix, are shown for
each of the regions defined by the clouds.\label{concentricFig}}
\end{center}
\end{figure}

To summarize, we can distinguish five concentric regions (four, if
$p-q \sim a_R$).  The effective magnetic charges seen within these are
\begin{equation}
    Q_{NA} = \cases{ {\rm diag}\, (0, 0, 0, 0) \, , & $r \gg p+q$ \cr\cr 
      {\rm diag}\, ( 0, -1, 0, 1) \, , & $p+q \gg  r \gg 
      {\rm Max}(p-q, a_R)$  \cr \cr
      {\rm diag}\, ( -1, -1, 1, 1) \, , & $p-q \gg r \gg a_R$ \cr \cr
      {\rm diag}\, ( -2, 0, 1, 1) \, , & $a_R \gg r \gg a_L$ \cr\cr
      {\rm diag}\, ( -2, 0, 0, 2) \, , & $a_L  \gg r$  \, . } 
\label{concentricCharges}
\end{equation} 
We illustrate this in Fig.~\ref{concentricFig}.

\section{Clouds and massless monopoles}
\label{candm}

With the explicit examples of the previous section in hand, we can now
examine more closely the relationship between the massless monopoles
and the non-Abelian clouds. The first thing to notice is that there
are six massless monopoles but only four non-Abelian clouds.  In other
words, the number of distinct clouds evident in the solutions is less
than the number of massless monopoles obtained by adding the
coefficients of the $\gamma_a$ in
Eq.~(\ref{quantizationSolution}).\footnote{This is not a completely
new phenomenon; the SU($N$) $(1,[1],\dots,[1],1)$ solutions have $N-3$
massless monopoles but only a single cloud.  However, these might have
been dismissed as special cases because they can all be obtained as
embeddings of $(1,[1],1)$ SU(4) solutions with a single massless
monopole.}
The number of degrees of freedom in the SU(6)
$(2,[2],[2],[2],2)$ solutions, 40, is precisely that expected when
there are a total of ten (four massive and six massless)
monopoles.  However, in all the examples we have studied, these degrees
of freedom parameterize the four massive monopoles and only four distinct,
though sometimes degenerate, clouds. In other words, there is not a
one-to-one correspondence between clouds and massless monopoles.
We expect that the solutions we
have studied are not exceptional in this regard and that a general
$(2,[2],[2],[2],2)$ will have four clouds.

To help understand this, recall that requiring that ${\bf
h}\cdot\bbeta_a$ is positive determines a unique set of simple roots
when the symmetry is maximally broken, but when the unbroken group
has a non-Abelian factor ${\bf h}\cdot\bbeta_a=0$ for some $\bbeta_a$'s and
so there are a number of possible choices for the simple roots, all
related by Weyl transformations.  To be specific, let us consider
SU(6).  Following Fig.~\ref{SUNCD}, we denote the simple roots for the
maximally broken case by $\bbeta_1$, $\bgamma_1$, $\bgamma_2$,
$\bgamma_3$, and $\bbeta_2$.  Each of these corresponds to a massive
fundamental monopole.  In the limit where the unbroken symmetry is
enlarged to U(1)$\times$SU(4)$\times$U(1), the $\bbeta_1$-monopole
becomes part of a multiplet of degenerate states transforming as a {\bf 4}
of SU(4); the remaining states in this multiplet correspond to the
roots $(\bbeta_1 + \bgamma_1)$, $(\bbeta_1 + \bgamma_1+ \bgamma_2)$,
and $(\bbeta_1 + \bgamma_1+ \bgamma_2+ \bgamma_3)$, all of which can
be obtained from $\bbeta_1$ by Weyl transformations.  Similarly, the
$\bbeta_2$-monopole is part of a $\bar {\bf 4}$ that also includes the
$(\bbeta_2 + \bgamma_3)$-, $(\bbeta_2 + \bgamma_3+ \bgamma_2)$-, and
$(\bbeta_2 + \bgamma_3+ \bgamma_2+ \bgamma_1)$-monopoles.  Clearly,
the straightforward correspondence of simple roots with elementary
monopoles and composite roots with multimonopoles has become more
complicated.

It is useful to make a detailed correspondence between the clouds and
the massless monopoles in specific solutions.  This can be done by
noting the effect of each cloud on the non-Abelian charge $Q_{NA}$.
Recall that the non-Abelian charges of the massive $\bbeta_1$- and
$\bbeta_2$-monopoles are
\begin{eqnarray} 
    Q_{NA}(\bbeta_1) &=& {\rm diag}\, (-1,0,0,0)   \cr
    Q_{NA}(\bbeta_2) &=& {\rm diag}\, (0,0,0,1) 
\end{eqnarray}
while those of the massless monopoles are
\begin{eqnarray}
  &&  Q_{NA}(\bgamma_1) = {\rm diag}\, (1,-1,0,0)   \cr
  &&  Q_{NA}(\bgamma_2) = {\rm diag}\, (0,1,-1,0)   \cr
  &&  Q_{NA}(\bgamma_3) = {\rm diag}\, (0,0,1,-1)   \cr
  &&  Q_{NA}(\bgamma_1+\bgamma_2) = {\rm diag}\, (1,0,-1,0)   \cr    
  &&  Q_{NA}(\bgamma_2+\bgamma_3) = {\rm diag}\, (0,1,0,-1)   \cr
  &&  Q_{NA}(\bgamma_1+ \bgamma_2+\bgamma_3) 
       = {\rm diag}\, (1,0,0,-1)  
\label{allmasslesscharges}
\end{eqnarray}
where  we have listed the charges for all of the massless positive
roots, not just the $\bgamma_a$.

As an example, consider the case of concentric Dancer clouds
illustrated in Fig.~\ref{concentricFig} and described by
Eq.~(\ref{concentricCharges}).  In the innermost region, $r \ll a_L$,
we have the field due to two $\bbeta_1$- and two $\bbeta_2$-monopoles.
Moving outward, we find a $\bgamma_3$-monopole at $r \approx a_L$, a
$\bgamma_1$-monopole at $r \approx a_R$, a
$(\bgamma_1+\bgamma_2)$-monopole at $r \approx p-q$ 
and finally a $(\bgamma_2+\bgamma_3)$-monopole
at $r \approx p+q$.

The fact that two of these roots are simple and two are composite is a
gauge-dependent statement; the sequence $\bgamma_3$, $\bgamma_1$,
$\bgamma_2$, $(\bgamma_1+\bgamma_2+\bgamma_3)$, for example, leads to
a physically equivalent configuration. 
However, the sum of these roots, $2\bgamma_1+2\bgamma_2+2\bgamma_3$,
(corresponding to a total of six massless monopoles) is invariant, as
is the fact that two of the roots are ``Dancer roots'' that are
orthogonal to one $\bbeta_a$ but not to the other.  We can also apply
a gauge transformation that replaces one of the $\bbeta_a$ by a
compound root.  Such a transformation will replace some of the
positive roots in the above sequences by a negative roots, but the sum
of the coefficients of the $\bgamma_a$ will remain unchanged.  The
correspondence between monopoles and clouds for various solutions is
indicated in Figs.~\ref{minoverFig}-\ref{concentricFig}.

Let us try to develop some rules to explain how and where the various
massless monopoles can appear in these solutions. It is useful to
begin by discussing what happens to the $\bgamma_a$-monopoles in the
maximally broken theory when the Higgs expectation value $\Phi_0$ is
varied.  As $\Phi_0$ approaches a value with an enlarged symmetry
group, the masses of some of the elementary bosons decrease,
eventually vanishing in the limit where the symmetry becomes
non-Abelian.  For an isolated $\bgamma_a$-monopole, the core radius is
inversely proportional to the corresponding elementary boson mass and
so grows monotonically and becomes infinite in the non-Abelian
limit. Similarly, for a configuration composed of two monopoles, a
$\bgamma_1$-monopole and a $\bgamma_2$-monopole, for example, the
component monopoles each grow without bound as the corresponding
masses vanish. When their core radii are much larger than the
separation of their centers, the two-monopole configuration is barely
distinguishable from a gauge-transformed one-monopole solution.

However, the situation can be different if the configuration contains
a monopole that remains massive in the non-Abelian limit. If the
fields in the massive core transform under the non-Abelian symmetry,
some of the massless gauge fields acquire an effective mass near the
massive core.  This can dramatically modify the behavior of the
massless monopole cores.  There are three possible situations.  One
possibility is that the massive and massless monopoles lie in mutually
commuting subgroups.  The massive monopole then has no effect on the
massless monopole, and the behavior described in the previous
paragraph is unchanged.  Another possibility is that the addition of
the massless monopole to the massive monopole corresponds to a gauge
transformation of the massive monopole, in which case no
gauge-invariant evidence of the massless monopole survives in the
non-Abelian limit.  The most interesting case is when neither of these
holds.  The massless monopole core then expands only until it reaches
the massive monopole, resulting in a cloud whose radius is equal to
the original intermonopole separation.\footnote{For a detailed study
of how the non-Abelian limit is approached in the last two cases, see
Ref.~\cite{Lu}.}

A similar pattern should be expected when several massive monopoles
are present.  A massless monopole with non-Abelian charge $q$ should
be able to form a finite size cloud enclosing a collection of
monopoles with total charge $Q_{NA}$ only if (1) $q$ does not lie in
the subgroup that leaves $Q_{NA}$ invariant and (2) $Q_{NA}' = Q_{NA}
+ q$ is not gauge-equivalent to $Q_{NA}$.  This is consistent with the
behavior of the clouds in both the SU(3) Dancer and the SU(4) $(1,[1],1)$
solutions.  In each case, combining the massless monopole with just
one of the massive monopoles would simply give a gauge-transformed
massive monopole and, indeed, we only find solutions where the cloud
encloses both of the massive monopoles.\footnote{The minimal cloud solutions
can be viewed as having the massless monopole exactly coincident with
one of the massive monopoles.  However, these should be understood as
a limiting case in which the ellipsoidal cloud enclosing the two
massive monopoles has degenerated into a line.}

Let us apply these rules to the concentric Dancer cloud solutions of
Sec.~\ref{concentricSubsec}.  We start with two $\bbeta_1$- and two
$\bbeta_2$-monopoles, with a total non-Abelian charge
\begin{equation}
   Q_1=  2 Q_{NA}(\bbeta_1) + 2 Q_{NA}(\bbeta_2)
             = {\rm diag}\, (-2,0,0,2)  \, ,
\label{firstQNA}
\end{equation}
and want to add one of the massless monopoles whose charges were
enumerated in Eq.~(\ref{allmasslesscharges}).  There are four
possibilities: we can add a massless monopole corresponding to:
\begin{enumerate}
  \item[a)] $\bgamma_1$ or $(\bgamma_1+\bgamma_2)$; these give 
gauge-equivalent solutions in the $Q_1$ background.
  \item[b)] $\bgamma_3$ or $(\bgamma_2+\bgamma_3)$; these are
     similarly gauge-equivalent.
    \item[c)] $\bgamma_2$.  
    \item[d)] $(\bgamma_1+\bgamma_2+\bgamma_3)$.
\end{enumerate}
Cases (a) and (b) differ only by the interchange of the left and right
Dancer solutions, and  need not be considered separately.  Case (c)
is ruled out because $\bgamma_2$ lies in the SU(2) subgroup that
leaves Eq.~(\ref{firstQNA}) invariant.  Thus, it is sufficient to
consider cases (b) and (d).  We start with the former, and add a
$\bgamma_3$-monopole to get
\begin{equation}
   Q_2=  Q_1 + Q_{NA}(\bgamma_3) 
             = {\rm diag}\, (-2,0,1,1)  \, .
\label{secondQNA}
\end{equation}
The possible choices for the next massless monopole are:
\begin{enumerate}
  \item[a)] $\bgamma_1$.
  \item[b)] $(\bgamma_1+\bgamma_2)$ or 
     $(\bgamma_1+\bgamma_2+\bgamma_3)$; these are gauge-equivalent 
     to each other.
  \item[c)] $\bgamma_3$; this is ruled out because it lies in 
      the SU(2) symmetry group of Eq.~(\ref{secondQNA}).
  \item[d)] $\bgamma_2$ or $(\bgamma_2+\bgamma_3)$; these 
      just gauge transform $Q_2$, [e.g., $Q_2 + Q_{NA}(\bgamma_2)
       = {\rm diag}\, (-2,1,0,1)$] and so cannot give rise to a cloud.
\end{enumerate}
Once more, we have two acceptable choices, (a) and (b).  If we take the
former, we obtain 
\begin{equation}
   Q_3=  Q_2 + Q_{NA}(\bgamma_1)
             = {\rm diag}\, (-1,-1,1,1)\, .
\label{thirdQNA}
\end{equation}

We cannot add either a $\bgamma_1$- or a $\bgamma_3$-monopole to this,
since both lie in the SU(2)$\times$SU(2) symmetry group of $Q_3$.  The
remaining four possibilities are all gauge-equivalent.  Choosing
$(\bgamma_1+\bgamma_2)$, we obtain
\begin{equation}
   Q_4=  Q_3 + Q_{NA}(\bgamma_1+\bgamma_2)
             = {\rm diag}\, (0,-1,0,1)\, .
\label{fourthQNA}
\end{equation}

At this point, the remaining massless monopoles only allow 
three possibilities: $\bgamma_2$, $\bgamma_3$, and
$(\bgamma_2+\bgamma_3)$.  The first two are excluded because they
give gauge transforms of $Q_4$.  The third gives
\begin{equation}
   Q_5=  Q_4 + Q_{NA}(\bgamma_2+\bgamma_3)
             = {\rm diag}\, (0,0,0,0)\, .
\end{equation}
We therefore recover the sequence $\bgamma_3$,
$\bgamma_1$, $(\bgamma_1+\bgamma_2)$, $(\bgamma_2+\bgamma_3)$ that we
found previously.

We now must return to the alternatives that we did not follow.  At the
first step, we could have taken case (d) and added a
$(\bgamma_1+\bgamma_2+\bgamma_3)$-monopole to $Q_1$, thus reaching
in one step a charge that was gauge-equivalent to $Q_4$.  
This can be viewed as simply a degenerate case of the four-step
sequence, corresponding to the situation where $a_L = a_R = p-q$.  The
other place where we had an option was in adding on to $Q_2$, where we
could have followed case (b) and added a
$(\bgamma_1+\bgamma_2)$-monopole.  Once again, this is simply a
degenerate case, equivalent to taking $a_R=p-q$.

Another instructive example is the solution with two minimal Dancer
clouds that was analyzed in Sec.~\ref{alignedMinimal} and illustrated
in Fig.~\ref{minoverFig}.  The two $\bbeta_1$-monopoles ``anchor'' a
massless monopole corresponding (by a gauge choice) to $(\bgamma_1 +
\bgamma_2)$.  In this solution the corresponding ellipsoidal Dancer
cloud has degenerated into a line, and the fields are those
appropriate to two massive monopoles, one with non-Abelian charge
\begin{equation}
    Q_{NA}(\bbeta_1) = \rm diag\,(-1,0,0,0) 
\end{equation} 
at ${\bf x}_1$, and one with 
\begin{equation}
    Q_{NA}(\bbeta_1+\bgamma_1 + \bgamma_2) = \rm diag\,(0,0,-1,0) 
\end{equation} 
at ${\bf x}_2$.\footnote{It seems obvious here that the $(\bgamma_1
+\bgamma_2)$-monopole is coincident with one of the massive monopoles.
Nevertheless, the position of this massless monopole, as defined by the
boundary values of $22$-components of the Nahm data, can, by an
appropriate change of basis, be taken to be any point along the line
joining the two $\bbeta_1$-monopoles.  A similar phenomenon occurs in
the minimal cloud $(1,[1],1)$ solutions in SU(4).}  
Similarly, a $(\bgamma_2 + \bgamma_3)$-monopole forms a minimal Dancer
cloud about the $\bbeta_2$-monopoles, effectively yielding a pair of
massive monopoles with charges $\rm diag\,(0,1,0,0)$ (at ${\bf x}_3$)
and $\rm diag\,(0,0,0,1)$ (at ${\bf x}_4$).  At this point we are left
with only two massless monopoles, a $\bgamma_1$ and a $\bgamma_3$.
The former can form a cloud anchored by the monopoles at ${\bf x}_1$ 
and ${\bf x}_3$, whose total charge is $\rm diag\,(-1,1,0,0) =
-Q_{NA}(\bgamma_1)$.  It could not, on the other hand, have formed a cloud
about the ${\bf x}_1$- and ${\bf x}_4$- or the ${\bf x}_2$- and ${\bf
x}_3$-monopoles, because in each case the effect of the
$\bgamma_1$-monopole would have only been to gauge transform the sum
of the charges of the two massive monopoles.  By similar arguments,
the $\bgamma_3$-monopole can only condense about the monopoles at 
${\bf x}_2$ and ${\bf x}_4$.

Similar analyses can be applied to the other limiting cases studied
in Sec.~\ref{twotwotwosolutions}.  In all cases, the appearance of the 
massless monopoles and their clouds is consistent with the rules
outlined above.

\section{Summary and concluding remarks}

Our goal in this paper has been to gain further insight into the
massless monopoles that can arise when a gauge theory is spontaneously
broken to a subgroup containing a non-Abelian factor.  These can be
viewed as limiting cases of massive fundamental monopoles of the
maximally broken theory, in the sense that they have the same number
of degrees of freedom and that the moduli space metric describing
their dynamics is a smooth limit of that for the maximally broken
case.  In contrast to the massive fundamental monopoles, however, they
cannot be realized as isolated classical solutions.  Instead, they are 
manifested through non-Abelian clouds surrounding one or more massive
monopoles.  We have focussed on the $(2,[2],\dots,[2],2)$ solutions of
SU($N$) broken to U(1)$\times$SU($N-2$)$\times$U(1).  These solutions
display a much richer structure than the one-cloud solutions that have
been previously studied.  The case of $N=6$ is generic, in the sense
that the solutions for $N>6$ can all be obtained by embedding the
SU(6) solution; we will discuss the restriction to smaller groups, and
in particular SU(4), below.

We have used the Nahm construction to obtain these solutions.  A
crucial ingredient in the application of this construction is the fact
that the Nahm data for the SU(6) problem are equivalent to two sets of
SU(3) Dancer Nahm data, together with a set of constant jump data.  In
the particular case where the jump data are large compared to the
other scales in the problem, they define a pair of nested ``SU(4)
clouds'' that enclose two independent Dancer solutions, characterized by
Dancer cloud parameters $a_L$ and $a_R$.

The spacetime fields of the SU(6) solution can be obtained from those
of the related Dancer solutions by purely algebraic manipulations.
Although the Dancer fields are not explicitly known for arbitrary
parameters, analytic approximations can be obtained both in the limit where
$a$ is much greater than the separation of the massive monopoles 
and in the case of minimal Dancer cloud.  Using these
approximations, we have obtained the leading behavior of the SU(6)
solutions for a number of limiting cases.

In these solutions, the clouds divide space into distinct regions, each
of which can be characterized by a non-Abelian magnetic charge
$Q_{NA}$.  By comparing the values of $Q_{NA}$ in adjacent regions,
one can establish a correspondence between a specific cloud and a
particular massless monopole.  In each case, one or more clouds are
associated with massless monopoles corresponding to compound, rather
than simple, roots.  As a result, the number of fundamental massless
monopoles (as defined, e.g., by the counting of degrees of freedom),
which is obtained by adding the coefficients of the simple roots in
Eq.~(\ref{quantizationSolution}), is greater than the number of
distinct clouds.  The SU(6) solutions contain six massless monopoles
but only four clouds, of which two are inherited from the Dancer
solutions and two are SU(4) clouds associated with the jump data.

Our explicit solutions also help clarify the role of the parameters
that enter the general $(2,[2],[2],[2],2)$ solution.  The moduli space
is 40-dimensional, with 17 of these corresponding to the unbroken
U(1)$\times$SU(4)$\times$U(1) symmetry.  The remaining 23 parameters
were enumerated at the end of Sec.~\ref{parametersection}.  Twelve of
these specify the positions of the massive monopoles, and two others
give the relative U(1) phase between pairs of identical massive
monopoles.  As with the U(1) phase parameter in the SU(2) two-monopole
solution, the effect of these phases falls exponentially with the
monopole separation.  They have, therefore, played no role in our
analysis, which focussed on the region outside of the massive monopole
cores.

The remaining nine parameters are associated with the non-Abelian
clouds.  Roughly speaking, four of these ($a_L$, $a_R$, $p$, and $q$)
are cloud size parameters and five determine the
relative SU(2) orientations of the Dancer and SU(4) clouds.  However,
as the example of Sec.~\ref{coincidentMinimal} shows, this division
between size and orientation parameters is not completely clearcut,
since the shape of the cloud can depend upon the orientation
parameters.

It is instructive to compare the SU(2) orientation parameters with the
U(1) phase parameters.  The latter would be true symmetry parameters,
having no effect on the form of the solutions, were it not for the
interactions between the U(1)-charged fields in the cores of the
massive monopoles.  The exponential falloff in the effect of these
parameters results from the fact that fields with U(1) charge are all
massive.  The role of the SU(2) parameters is similar, but because
there are massless fields carrying SU(2) charges, the effect of these
parameters only falls as a power of distance.  Note that the distance to
the SU(4) cloud is crucial here, since the Dancer solutions lie in
mutually commuting subgroups and therefore have no direct
interactions.

It is also useful to consider the application of our results to
smaller groups.  In particular, let us try to identify those SU(6)
solutions that can be viewed as embeddings of the $(2,[2],2)$ SU(4)
solutions.  In the Nahm construction, these are distinguished by the
fact that their jump data includes only two independent $a_p$, rather
than four.  In terms of the construction of Sec.~\ref{KKKsolutions},
this translates into the statement that two of the eigenvalues of $K$
vanish. A more physical characterization is based on the fact that the
SU(4) solutions contain only two massless monopoles, and thus, at most,
two distinct clouds.  Hence, they must correspond to SU(6) solutions
in which some of the clouds are coincident.

Let us apply these considerations to some of the examples discussed in
Sec.~\ref{twotwotwosolutions}:

1. {\it Minimal Dancer clouds with all SU(2)'s aligned}
   (Sec.~\ref{alignedMinimal}): There are two ways for $K$ to have two
   vanishing eigenvalues.  One is to set $p+q = |{\bf x}_1 - {\bf
   x}_3|$ and $p-q = |{\bf x}_2 - {\bf x}_4|$.  This is equivalent to
   a pair of SU(4) $(1,[1],1)$, each of which has a minimal cloud and
   is thus an SU(3) solution embedded in SU(4).
   The alternative is to take $p=q$ and ${\bf x}_2 - {\bf
   x}_4$.   This is equivalent to an SU(4) $(1,[1],1)$ solution, with
   massive monopoles at ${\bf x}_1$ and ${\bf x}_3$ and cloud size
   $p+q=2p$, together with a minimal cloud SU(4) $(1,[1],1)$ solution
   with coincident massive monopoles at ${\bf x}_2$.

2. {\it Minimal Dancer clouds with coincident massive monopoles}
   (Sec.~\ref{coincidentMinimal}): Equation~(\ref{coincidentT}) shows
   that two eigenvalues of $K$ vanish if $p=q$.
   Equation~(\ref{coincidentB}) then implies that the nonvanishing
   components of the Higgs field must lie in the subspace orthogonal
   to $M^-$.  If the angle $\alpha$ defined in Eq.~(\ref{cosAlphaDef})
   vanishes, this gives the second solution of case (1), but with the
   restriction that ${\bf x}_1 ={\bf x}_3$.  On the other hand, if
   $\alpha \ne 0$, corresponding to misaligned SU(2) orientations,
   then the outermost cloud (specified roughly by $\lambda_+ = 1$)
   encloses all four of the massive monopoles.

3. {\it Two widely separated large Dancer clouds}
   (Sec.~\ref{wideSepSubsec}): The vanishing of two of the eigenvalues
   in Eq.~(\ref{widelySepEvalues}) implies that $p = R +O(a)$ and $q =
   O(a)$; a more careful analysis of Eq.~(\ref{KforTwoLargeD}) is
   needed to determine the exact values.  This leads to a somewhat
   degenerate version of the situation depicted in
   Fig.~\ref{largeRFig}, with the two Dancer clouds and the two SU(4)
   clouds not clearly distinguishable.

4. {\it Two concentric large Dancer clouds}
   (Sec.~\ref{concentricSubsec}): If we insist that $a_R \gg a_L$, as
   in our previous analysis, then at most one eigenvalue of $K$ can
   vanish, implying that this cannot be reduced to an SU(4) solution.
   On the other hand, examination of Eq.~(\ref{KforTwoLargeD})
   suggests that there should be embedded SU(4) solutions with $a_R
   \approx a_L \approx p-q$.  In terms of the discussion following
   Eq.~(\ref{fourthQNA}), this corresponds to having the clouds from
   the $\bgamma_1$-, the $\bgamma_3$-, and the
   $(\bgamma_1+\bgamma_2)$-monopoles all coincident, with the fourth,
   $(\bgamma_2+\bgamma_3)$, cloud at a (possibly much larger) radius
   $p+q$.

For SU(4) broken to U(1)$\times$SU(2)$\times$U(1), a
solution containing $k_1$ $\bbeta_1$-monopoles and $k_2$
$\bbeta_2$-monopoles has vanishing SU(2) magnetic charge, and thus no
asymptotic non-Abelian Coulomb field, if there are exactly
$(k_1+k_2)/2$ massless $\bgamma$-monopoles.  However, this tells us
little about the magnetic field at finite distances.  In particular,
we would like to know whether the massive monopoles are ``paired up''
to form SU(2)-neutral combinations composed of two massive monopoles
enclosed by a single massless monopole cloud.  It is certainly clear
that configurations of this sort must exist.  Thus, there should be
$(2,[2],2)$ solutions that are approximately the sum of two widely
separated $(1,[1],1)$ solutions, as well as others that correspond to
two widely separated Dancer solutions, one formed from the
$\bbeta_1$-monopoles and one from the $\bbeta_2$-monopoles.  The
question is whether this is the generic situation.  The solutions
described in paragraph (2) are instructive in this regard.  For $\alpha=0$
these do, indeed, naturally separate into a pair of $(1,[1],1)$
solutions.  However, for the presumably more generic case of nonzero
$\alpha$, we find that the ``$\lambda_+$-cloud'' encloses all four
massive monopoles.  This strongly suggests that in general there is no
clear pairing of massive monopoles.

Finally, we want to indicate briefly two areas for future research.
The first is the determination of the $(2,[2],\dots,[2],2)$ moduli
space metric, which determines the low-energy dynamics of the
monopoles.  In general, the explicit determination of such metrics is
a difficult task.  However, one might hope that relation between the
Dancer Nahm data and the $(2,[2],\dots,[2],2)$ Nahm data could be
exploited to allow a construction of the moduli space metric in terms
of the Dancer metric \cite{D1}.  Secondly, one would like to understand
better the quantum theory of monopoles with non-Abelian charges, and
in particular the complex interplay of the non-Abelian electric and 
magnetic charges \cite{bais}.  The knowledge of classical multicloud 
solutions that we have gained should help provide a starting point for 
this investigation.

\acknowledgments This work was supported in part by the
U.S. Department of Energy. CJH is grateful to the Fulbright Commission
and to the Royal Commission for the Exhibition of 1851 for financial
support and to the Department of Physics, Columbia University for
hospitality.


\thebibliography{99}

\bibitem{montonen}
C.~Montonen and D.~I.~Olive,
Phys.\ Lett.\ B {\bf 72}, 117 (1977).

\bibitem{Lee:1996vz}
K.~Lee, E.~J.~Weinberg, and P.~Yi,
Phys.\ Rev.\ D {\bf 54}, 6351 (1996).

\bibitem{Bogomolny:1975de}
E.~B.~Bogomolny,
Sov.\ J.\ Nucl.\ Phys.\  {\bf 24}, 449 (1976)
[Yad.\ Fiz.\  {\bf 24}, 861 (1976)].

\bibitem{Prasad:kr} 
M.~K.~Prasad and C.~M.~Sommerfield,
Phys.\ Rev.\ Lett.\  {\bf 35}, 760 (1975).

\bibitem{quantizationRef}
F.~Englert and P.~Windey,
Phys.\ Rev.\ D {\bf 14}, 2728 (1976);
P.~Goddard, J.~Nuyts, and D.~I.~Olive,
Nucl.\ Phys.\ B {\bf 125}, 1 (1977).

\bibitem{Weinberg:1979zt}
E.~J.~Weinberg,
Nucl.\ Phys.\ B {\bf 167}, 500 (1980).

\bibitem{pathologies}
A.~Abouelsaood,
Phys.\ Lett.\ B {\bf 125}, 467 (1983);
P.~Nelson and A.~Manohar,
Phys.\ Rev.\ Lett.\  {\bf 50}, 943 (1983);
A.~P.~Balachandran, G.~Marmo, N.~Mukunda, J.~S.~Nilsson,
E.~C.~Sudarshan, and F.~Zaccaria, 
Phys.\ Rev.\ Lett.\  {\bf 50}, 1553 (1983);
P.~Nelson and S.~R.~Coleman,
Nucl.\ Phys.\ B {\bf 237}, 1 (1984).

\bibitem{Weinberg:ev}
E.~J.~Weinberg,
Nucl.\ Phys.\ B {\bf 203}, 445 (1982).

\bibitem{Weinberg:jh}
E.~J.~Weinberg,
Phys.\ Lett.\ B {\bf 119}, 151 (1982).

\bibitem{WY} 
E.~J.~Weinberg and P.~Yi,
Phys.\ Rev.\ D {\bf 58}, 046001 (1998).

\bibitem{D1} 
A.~S.~Dancer,
Commun.\ Math.\ Phys.\  {\bf 158}, 545 (1993).

\bibitem{Dancer:kj}
A.~S.~Dancer,
Nonlinearity {\bf 5}, 1355 (1992).  

\bibitem{Lee:1997ny}
K.~Lee and C.~Lu,
Phys.\ Rev.\ D {\bf 57}, 5260 (1998).

\bibitem{Ho} 
C.~J.~Houghton,
Phys.\ Rev.\ D {\bf 56}, 1220 (1997).

\bibitem{HIM} 
C.~J.~Houghton, P.~W.~Irwin, and A.~J.~Mountain,
JHEP {\bf 9904}, 029 (1999).

\bibitem{Nahmconst} 
W. Nahm, in {\it Monopoles in quantum field theory}, Craigie et
al. eds.  (World Scientific, Singapore, 1982); 
in {\it Gauge theories and lepton hadron interactions},
edited by Z. Horvath et al. (Central Research Institute for Physics, Budapest,
1982); 
in {\it Structural Elements in Particle Physics and Statistical
Mechanics}, edited by J. Honerkamp et al. (Plenum, New York, 1983) 
in {\it Group theoretical methods in physics}, Denardo et
al. eds. (Springer-Verlag, 1984). 

\bibitem{ADHM} 
M.~F.~Atiyah, N.~J.~Hitchin, V.~G.~Drinfeld, and Y.~I.~Manin,
Phys.\ Lett.\ A {\bf 65}, 185 (1978).

\bibitem{H1} N.~J.~Hitchin,
Commun.\ Math.\ Phys.\  {\bf 89}, 145 (1983).

\bibitem{CG} 
E.~Corrigan and P.~Goddard,
Annals Phys.\  {\bf 154}, 253 (1984).

\bibitem{N1} 
W.~Nahm, 
Phys.\ Lett.\ B {\bf 90}, 413 (1980).

\bibitem{HM} 
J.~Hurtubise and M.~K.~Murray,
Commun.\ Math.\ Phys.\  {\bf 122}, 35 (1989).

\bibitem{DL} 
A.~S.~Dancer and R.~A.~Leese,
Phys.\ Lett.\ B {\bf 390}, 252 (1997).

\bibitem{I} 
P.~Irwin,
Phys.\ Rev.\ D {\bf 56}, 5200 (1997).

\bibitem{DL2} 
A.~S.~Dancer and R.~A.~Leese,
Proc. R. Soc. Lond. A {\bf 440}, 421 (1993).

\bibitem{Lu}
C.~Lu,
Phys.\ Rev.\ D {\bf 58}, 125010 (1998).

\bibitem{bais}
F.~A.~Bais and B.~J.~Schroers,
Nucl.\ Phys.\ B {\bf 512}, 250 (1998); 
B.~J.~Schroers and F.~A.~Bais,
Nucl.\ Phys.\ B {\bf 535}, 197 (1998).
\end{document}